\documentclass[%
onecolumn,%
oneside,%
floats,%
aps,%
 prd,%
nobibnotes,%
nofootinbib,%
amsmath,%
amssymb,%
amsfonts,%
amscd,%
  superscriptaddress,%
eqsecnum%
]{revtex4}

\usepackage[utf8]{inputenc}
\usepackage{graphicx,array,dcolumn}
\usepackage{cases}
\usepackage[paperwidth=210mm,paperheight=297mm,centering,hmargin=1.8cm,vmargin=2.5cm]{geometry}
\usepackage{enumerate}

\usepackage{hyperref}
\hypersetup{hidelinks}

\newcommand{\rar}{\rightarrow}
\newcommand{\lrar}{\leftrightarrow}

\def\mc{\mathcal}
\def\be{\begin{eqnarray}}
\def\ee{\end{eqnarray}}

\def\D{\mc D}
\def\-g{\sqrt{-g}}

\renewcommand\rho{\varrho}

\begin{document}

\title{General properties and kinetics of spontaneous baryogenesis }

\author{E.V. Arbuzova}
\email{arbuzova@uni-dubna.ru}
\affiliation{Novosibirsk State University, Novosibirsk, 630090, Russia}
\affiliation{Department of Higher Mathematics, Dubna State University, 141980 Dubna, Russia}

\author{A.D. Dolgov}
\email{dolgov@fe.infn.it}
\affiliation{Novosibirsk State University, Novosibirsk, 630090, Russia}
\affiliation{ITEP, Bol. Cheremushkinsaya ul., 25, 117259 Moscow, Russia}
\affiliation{Dipartimento di Fisica e Scienze della Terra, Universit\`a degli Studi di Ferrara\\
Polo Scientifico e Tecnologico - Edificio C, Via Saragat 1, 44122 Ferrara, Italy}

\author{V.A. Novikov}
\email{novikov@itep.ru}
\affiliation{ITEP, Bol. Cheremushkinsaya ul., 25, 117259 Moscow, Russia}

\begin{abstract}

General features of spontaneous baryogenesis are studied. The relation between the time derivative of the 
(pseudo)goldstone field and the baryonic chemical potential is revisited. It is shown that this relation essentially 
depends upon the representation chosen for the fermionic fields with non-zero baryonic number (quarks). 
The calculations of the cosmological baryon asymmetry are based on the kinetic equation generalized to the 
case of non-stationary background. The effects of the finite interval of the integration over time are also included 
into consideration. All these effects combined lead to a noticeable deviation of the magnitude of the baryon 
asymmetry from the canonical results.

 \end{abstract}

\maketitle

\section{Introduction \label{s-intro}}

The usual approach to cosmological baryogenesis is based on three well known Sakharov's conditions~\cite{ads}:
a)~nonconservation of baryonic number; b) breaking of C and CP invariance; c) deviation from thermal equilibrium.
There are however some interesting scenarios of baryogenesis for which one or several of the above conditions are not fulfilled.
A very popular scenario is the so called  spontaneous baryogenesis (SBG) proposed in 
refs~\cite{spont-BG-1,spont-BG-2,spont-BG-3}, for reviews see e.g.~\cite{BG-rev,AD-30}. 
The term "spontaneous" is related to spontaneous breaking of underlying symmetry of the theory.
It is assumed that in the unbroken phase the Lagrangian is
invariant with respect to the global $U(1)$-symmetry, which ensures conservation of the total baryonic number
including that of the Higgs-like field, $\Phi$, and the matter fields (quarks).
This symmetry is supposed to be spontaneously broken 
and in the broken phase the Lagrangian density  acquires the term
\be
{\cal L}_{SB} =  (\partial_{\mu} \theta) J^{\mu}_B\, ,
 \label{L-SB}
 \ee
 where $\theta$ is Goldstone field, or in other words, the phase of the field $\Phi$ 
 and $J^{\mu}_B$ is the baryonic current of matter fields (quarks). 
 Depending upon the form of the interaction of $\Phi$ with the matter fields,
 the spontaneous symmetry breaking (SSB) may lead to nonconservation of the baryonic  current of matter.
 If this is not so and
 $J^{\mu}_B$ is conserved, then integrating by parts eq. ~(\ref{L-SB}) we obtain a vanishing expression and hence
 the interaction~(\ref{L-SB}) is unobservable. 
 
 The next step  in the implementation of the SBG scenario
 is the conjecture that the Hamiltonian density corresponding to ${\cal L}_{SB}$ is 
 simply the Lagrangian density taken with the opposite sign:
 \be 
{\cal H}_{SB} = - {\cal L}_{SB} = - (\partial_{\mu} \theta) J^{\mu}_B\, .
\label{H-SB}  
\ee   
This could be true, however,  if the Lagrangian depended only on the field variables but not on their derivatives, as it is argued below.

For the time being we neglect the complications related to the questionable identification (\ref{H-SB}) and proceed further
in description of the SBG logic.

For the spatially homogeneous field $\theta = \theta (t)$ the Hamiltonian (\ref{H-SB})
is reduced to ${\cal H}_{SB} =  - \dot \theta\, n_B$, where $n_B\equiv J^4_B$ 
is the baryonic number density of matter, so it is tempting to identify $\dot \theta$ with the chemical potential, $\mu$, of the corresponding 
system, see e.g.~\cite{LL-stat}.
If this is the case, then in thermal equilibrium with respect to  the baryon non-conserving interaction the baryon asymmetry would evolve to:
\be 
n_B =\frac{g_S B_Q}{6} \left({\mu\, T^2} + \frac{\mu^3}{ \pi^2}\right) \rar
\frac{g_S B_Q}{6} \left({\dot \theta\, T^2} + \frac{\dot \theta ^3}{ \pi^2}\right)\,,
\label{n-B-of-theta}
\ee
where  $T$ is the cosmological plasma temperature, $g_S$ and $B_Q$ are respectively the number of the spin states 
and the baryonic number of quarks, which are supposed to be the bearers of the baryonic number.

It is interesting that for successful SBG two of the Sakharov's conditions for the generation of the cosmological 
baryon asymmetry, namely, breaking of thermal equilibrium and a violation of C and CP symmetries are 
unnecessary.  This scenario is analogous the baryogenesis in absence of CPT invariance, if the masses of particles and antiparticles are
different. In the latter case the generation of the cosmological baryon asymmetry can also proceed in thermal equilibrium~\cite{ad-yab-cpt,ad-cpt}.
In the SBG scenario the role of CPT "breaker" plays the external field $\theta (t)$.

However, in contrast with the usual saying, the identification $\dot\theta = \mu_B$ is incorrect. Indeed, 
if  $\dot \theta (t)$ is constant or slowly varying, then according to eq.~(\ref{H-SB}) it
shifts the energies of baryons with respect to antibaryons at the same spatial momentum, by $\dot\theta$. Thus
there would be different number densities of baryons and antibaryons in the plasma even if the corresponding chemical potential
vanishes. In this case the baryon asymmetry is determined by effective chemical potential $\mu_{eff} = \mu - \dot\theta$ to be
substituted into eq.~(\ref{n-B-of-theta}) instead of $\mu$. The detailed arguments are presented in sec.~\ref{s-kin-eq-canon}. It is also
shown there that the baryonic chemical potential tends to zero when the system evolves 
to the thermal equilibrium state. So in equilibrium the baryon asymmetry would be non-zero with vanishing chemical potential.

The picture becomes different if we use another representation for the quark fields.
Redefining the quark fields by the phase transformation, $ Q \rar \exp (i\theta/3 ) Q$, 
we can eliminate the term (\ref{L-SB}) from the Lagrangian,
but instead it would appear in the interaction term which violates B-conservation, see eq.~(\ref{L-of-theta-1}). Clearly in this case $\dot\theta$
is not simply connected to  the chemical potential. However, as is shown in the presented paper, 
the baryonic chemical potential in this formulation of the
theory would tend in equilibrium to $c\, \dot\theta$ with a constant coefficient $c$.
Anyway, as we see from the solution of the kinetic equation presented below, the physically meaningful expression of the
baryon asymmetry, $n_B$, expressed through $\theta$, is the same independently on the mentioned above two different 
formulations of the theory, though the values of the chemical potentials are quite different. Seemingly this difference
is related to non-accurate transition from the Lagrangian ${\cal L}_{SB}$ to the Hamiltonian ${\cal H}_{SB}$, made according 
to Eq.~(\ref{H-SB}). Such identification is true if the Lagrangian does not depend on the time derivative of the corresponding field,
$\theta (t)$ in the case under scrutiny. The related criticism of spontaneous baryogenesis can be found
in Ref.~\cite{AD-KF}, see also the review~\cite{AD-30}. 

Recently the gravitational baryogenesis scenario was suggested~\cite{gravBG-1}, see also \cite{gravBG-papers}. 
In these works the original SSB model was modified 
by the substitution of curvature scalar $R$ instead of the goldstone field $\theta $. With an advent of the $F(R)$-theories 
of modified gravity the gravitational baryogenesis was studied in their frameworks~\cite{gravBG-F-of-R} as well.   

In this paper the classical version of SBG is studied. We present an accurate derivation of the Hamiltonian for the Lagrangian which
depends upon the field derivatives. For a constant $\dot\theta$ and sufficiently large interval of the integration over
time the results are essentially the same as obtained in the previous considerations.
With the account of the finite time effects, which effectively break the energy conservation, the outcome of SBG becomes 
significantly different. We have also
considered an impact of a nonlinear  time evolution of the Goldstone field: 
\be
\theta = \dot\theta_0 t + \ddot\theta_0 \, t^2/2
\label{theta-Taylor2}
\ee
and have found that there can be significant deviations from the standard scenario with $ \dot\theta \approx const$.

A strong deviation from the standard results is also found  for the pseudgoldstone field oscillating near the minimum of the
potential $U(\theta)$.

The paper is organized as follows.  In section~\ref {s-ssb} the general features of the spontaneous breaking of baryonic $U(1)$-symmetry 
are described and the (pseudo)Goldstone  mode, its equation of motion, as well as the equations of motion of the quarks are introduced. 
In sec.~\ref{s-H-v-L} the construction of the Hamiltonian density from known Lagrangian is considered. Next, in sec.~\ref{s-kin-eq-canon} the standard kinetic 
equation in stationary background is presented. Sec.~\ref{s-evol-theta-G} is devoted to the generation of cosmological baryon asymmetry 
with out-of-equilibrium
purely Goldstone field. The pseudogoldstone case is studied in sec.~\ref{s-psevdo}. In sec.~\ref{s-kin-eq} we derive kinetic equation
in time dependent external field and/or for the case when energy is not conserved because of finite limits of integration over time. Several
examples, when such kinetic equation is relevant, are presented in sec.~\ref{s-examples}. Lastly in sec.~\ref{s-conclude} we conclude.

\section{Spontaneous symmetry breaking and goldstone mode \label{s-ssb}}

We start with the theory of a complex scalar field $\Phi$ interacting with fermions $Q$ and  $L$ 
with the Lagrangian:
\be
{\cal L }(\Phi) =  g^{\mu\nu} \partial_\mu \Phi^*
\partial_\nu \Phi - V(\Phi^* \Phi) + \bar Q (i \gamma^\mu \partial_\mu - m_Q)\,Q 
+  \bar L ( i \gamma^\mu \partial_\mu - m_L) L + {\cal L}_{int}(\Phi, Q, L)\, ,
\label{S-Phi}
\ee
where it is assumed that $Q$ and $\Phi$ have nonzero baryonic numbers, while $L$ have not. Here $V(\Phi^* \Phi)$ is
the self-interaction potential of $\Phi$ defined below in Eq.~(\ref{V-of-Phi}).
The interaction Lagrangian
${\cal L}_{int} $ describes the coupling between $\Phi $ and fermionic fields. In the toy model studied
below we take it in the form:
\be
{\cal L}_{int} =  \frac{\sqrt 2}{m_X^2} \frac{\Phi}{f}\, (\bar L \gamma_{\mu} Q )(\bar Q^c  \gamma_{\mu} Q) +
h.c. \, , 
\label{L-int}
\ee
where $Q^c$ is charged conjugated quark spinor and $m_X$ and $f$ are parameter with dimension of mass.
We prescribe to $\Phi$ and $Q$ the baryonic numbers $(-1)$ and $1/3$ respectively, so the interaction (\ref{L-int})
conserves the baryonic number.
The interaction of this type can appear e.g. in $SU(5)$ Grand Unified Theory.  For simplicity, in our toy model  we do not take 
into account the quark colors.

$Q$ and $L$ can be any fermions, not necessarily
quarks and leptons of the standard model.  They can
be e.g. new heavy fermions possessing similar or the same quantum numbers as the quarks and leptons of the standard model.
They should be coupled to the ordinary quarks and leptons in such a way that the baryon asymmetry in the Q-sector would be 
transformed into the asymmetry of the observed baryons.

Other forms of ${\cal L}_{int} $ can be considered leading e.g. to transition 
$3L \lrar Q$ or $2 Q \lrar 2\bar Q$.  They are not permitted for the standard quarks. 
However, for the usual quarks the process $3 q \lrar 3\bar q$ is permitted.  Note that the kinetics
of all these processes is similar. 
We denote by $q$ the usual quarks or the fermionic field with the same quantum numbers.

The Lagrangian (\ref{S-Phi}) is invariant under the following $U(1)$ transformations with constant $\alpha$:
\be
\Phi \rightarrow e^{i \alpha} \Phi, ~~~~Q \rightarrow e^{- i \alpha/3} Q,
{}~~~~L \rightarrow L \, .
\label{phase}
\ee
In the unbroken symmetry phase this invariance leads to the conservation of the total baryonic number of
$\Phi $ and of quarks.
In realistic model the interaction of left- and right-handed fermions may be different but 
we neglect this possible difference in what follows.

The global $U(1)$-symmetry is assumed to be spontaneously broken at the
energy scale $f$  via the potential of the form:
\be
V(\Phi^* \Phi) = \lambda \left(\Phi^* \Phi - f^2/2 \right)^2  .
\label{V-of-Phi}
\ee
This potential reaches minimum at the vacuum expectation value of $\Phi $ equal to
$\langle \Phi \rangle = f e^{i\phi_0/f}/\sqrt{2}$ with an arbitrary constant phase $\phi_0$.

Below scale $f$ we can neglect the heavy radial mode of
$\Phi$ with the mass $m_{radial} = \lambda^{1/2} f$, since being very
massive it is frozen out, but this simplification is not necessary and is not essential for the baryogenesis.
The remaining light degree of freedom is the variable field $\phi$, which is the Goldstone boson of the
spontaneously broken $U(1)$. Up to a constant factor the field $\phi$  
is the  angle around the bottom of the Mexican hat potential given by eq.~(\ref{V-of-Phi}). 
Correspondingly we introduce the dimensionless angular field $\theta \equiv \phi/f$
and thus $\Phi = \langle \Phi \rangle \exp (i \theta )$.

The low energy limit of the Lagrangian (\ref{S-Phi}) in the broken phase, which effectively describes the dynamics of $\theta$-field, 
has the form: 
\be
{\cal L}_1 (\theta)
= {f^2\over 2} \partial_\mu \theta \partial^{\mu} \theta + \bar Q_1
(i \gamma^{\mu} \partial_\mu - m_Q) Q_1 +  \bar L (
i \gamma^{\mu} \partial_\mu - m_L) L +
 \left(\frac{e^{i \theta }}{ m_X^2} \, (\bar L \gamma_{\mu} Q_1 )(\bar Q_1^c  \gamma_{\mu} Q_1) +
  h.c.\right) - 
 U(\theta)\, .
 \label{L-of-theta-1}
 \ee
Here we added the  potential  $ U(\theta) $, which may be induced by an  explicit symmetry 
breaking and can lead, in particular, to a nonzero mass of $\theta$. We use the notation $Q_1$ for the quark field to 
distinguish it from the phase rotated field $Q_2$ introduced below in Eq.~ (\ref{L-of-theta-2}). 
In a realistic model the quark fields should be (anti)symmetrized 
 with respect to color indices, omitted here for simplicity. 
 
 If $U(\theta ) = 0$, the theory remains invariant with respect to the global 
$U(1)$-transformations (i.e. the transformations with a constant phase $\alpha$):
 \be
Q \rightarrow e^{ - i \alpha/3} Q, ~~~~L \rightarrow L,
{}~~~~\theta \rightarrow \theta + \alpha  \, .
\label{global-trans}
\ee
The phase transformation of the quark field  
with the coordinate dependent phase $\alpha =  \theta (t, \bf x)$ introduces the new field $Q_1 = e^{- i \theta /3} Q_2$. In terms of this 
field the  Lagrangian~(\ref{L-of-theta-1}) turns  into:
\be
{\cal L}_2 (\theta)
= {f^2\over 2} \partial_\mu \theta \partial^{\mu} \theta + \bar Q_2
(i \gamma^{\mu} \partial_\mu - m_Q) Q_2 +  \bar L (
i \gamma^{\mu} \partial_\mu - m_L) L + \nonumber \\
 \left(\frac{1}{ m_X^2} \, (\bar Q_2 \gamma_{\mu} L )(\bar Q_2  \gamma_{\mu} Q_2^c) +
  h.c.\right) + (\partial_\mu \theta) J^\mu - U(\theta) \, , 
  \label{L-of-theta-2}
 \ee
where the quark baryonic current is $J_\mu =(1/3) \bar Q \gamma_\mu Q$. Note that the form of this current is the same  
in terms of $Q_1$ and $Q_2$.

The equation of motion for the quark field $Q_1$ which follows from  Lagrangian (\ref{L-of-theta-1}) has the form:
\be
(i \gamma^{\mu} \partial_\mu - m_Q) Q_1 +  \frac{e^{-i \theta }}{ m_X^2} \left[  \gamma_{\mu} L (\bar Q_1  \gamma_{\mu} Q_1^c) +
2 \gamma_{\mu} Q_1^c (\bar Q_1 \gamma_{\mu} L )  \right]  =0\,.
  \label{dirac-1}
 \ee
Analogously the equation of motion for the phase rotated field $Q_2$ derived from Lagrangian (\ref{L-of-theta-2})  is 
\be
\left(i \gamma^{\mu} \partial_\mu - m_Q + \frac{1}{3}\,\gamma ^\mu \partial_\mu \theta \right) Q_2 
+  \frac{1}{ m_X^2} \left[  \gamma_{\mu} L (\bar Q_2  \gamma_{\mu} Q_2^c) +
2 \gamma_{\mu} Q_2^c (\bar Q_2 \gamma_{\mu} L )  \right]  =0\,.
  \label{dirac-2} 
 \ee
Equations for $\theta$-field derived from these two Lagrangians in flat space-time have respectively the forms:
\be 
f^2 (\partial _t^2 - \Delta ) \theta + U'(\theta) 
+ \left[ \frac{i \,e^{- i \theta }}{ m_X^2} \, (\bar Q_1 \gamma_{\mu} L )(\bar Q_1  \gamma_{\mu} Q_1^c) +
  h.c.\right] = 0 
  \label{d2-theta-1}
  \ee 
  and
  \be 
f^2 (\partial _t^2 - \Delta ) \theta + U'(\theta)  +  \partial _\mu J^\mu_B = 0\, ,
   \label{d2-theta-2}
  \ee 
  where $U'(\theta) = dU/d\theta $. 
  
 Using either the equation of motion (\ref{dirac-1}) or (\ref{dirac-2}) we can check that the baryonic
 current is not conserved. Indeed, its divergence is:
 \be
 \partial_\mu J^\mu_B = \frac{i \,e^{-i\theta}}{m_X^2} (\bar Q_1 \gamma_\mu Q_1^c) (\bar Q_1 \gamma^\mu L) 
 + h.c.
 \label{dmu-Jmu}
 \ee 
 The current divergence in terms of the "rotated" field $Q_2$ has the same form
 but without the factor $\exp(-i\theta)$.
 So the equations of motion for $\theta$ in both cases (\ref{d2-theta-1}) and (\ref{d2-theta-2}) coincide, as expected. 

Eq.~(\ref{d2-theta-2}) expresses the law of the total baryonic current conservation in the unbroken phase. When the symmetry
is broken, the non-conservation of the physical baryons (in our case of "quarks") becomes essential and may lead to the observed 
cosmological baryon asymmetry.
Such B-non conserving  interaction may have many different forms. The one presented above describes transition of three quark-type fermions into 
(anti)lepton. There may be transformation of two or three quarks into equal number of antiquarks. Such interaction describes neutron-antineutron
oscillations, now actively looked for~\cite{n-bar-n}. There even can be a "quark" transition into three "leptons". Depending on the interaction type
the relation between $\dot\theta$ and the effective chemical potential would have different forms, i.e. different values of the proportionality
coefficient $c$ mentioned in the Introduction.

In the spatially homogeneous case, when $\partial_\mu J^\mu_B = \dot n_B $ and
$\theta = \theta (t)$, and if $U(\theta) = 0$,  equation (\ref{d2-theta-2}) can be easily integrated giving:
\be
f^2 \left[\dot \theta (t) - \dot \theta (t_{in})\right] =  -n_B (t) +  n_B (t_{in})  \,.
\label{nB-of-t}
\ee
It is usually assumed that the initial baryon asymmetry vanishes, $n(t_{in}) = 0$. 

The evolution of $n_B (t)$ is governed by the kinetic equation discussed in Sec. \ref{s-kin-eq-canon}. This equation allows to express $n_B$ through $\theta (t)$
and to obtain the closed systems of, generally speaking, integro-differential equations. In thermal equilibrium the relation between
$\dot\theta$ and $ n_B $ may become an algebraic one, but this is true only in the case when the interval of the integration
over time is sufficiently long and
 if $\dot\theta$ is constant or slowly varying  function of  time.

In the cosmological Friedmann-Robertson-Walker (FRW) background the equation of motion of $\theta $ (\ref{d2-theta-2}) becomes:
  \be
  f^2 (\partial _t + 3 H  ) \dot\theta - a^{-2}(t)\, \Delta \theta 
    + U'(\theta)   =  - (\partial _t + 3 H ) n_B, 
  \label{d2-theta-FRW-1}
  \ee 
  where $a(t)$ is the cosmological scale factor and $H=\dot a/a$ is the Hubble parameter. For the homogeneous theta-field, 
  $\theta = \theta (t)$, this equation turns into:  
   \be
  f^2 (\partial _t + 3 H  ) \dot\theta 
  + U'(\theta)   =  - (\partial _t + 3 H ) n_B.
  \label{d2-theta-FRW}
  \ee  
  
We do not include the curvature effects in the Dirac equations because they are not essential for what follows. Still we have
taken into account the  impact of the cosmological expansion on the current divergence using the covariant derivative 
in the FRW space-time: 
 $\D_\mu J^\mu =\dot n_B+ 3 Hn_B$.

\section{Hamiltonians versus Lagrangians} \label{s-H-v-L}

Though, as we see in secs.~\ref{s-kin-eq-canon} and \ref{s-kin-eq}, the baryon asymmetry originated in the frameworks of SBG is proportional to
$\dot \theta$ in many interesting cases, as justly envisaged in refs.~\cite{spont-BG-1,spont-BG-2}, 
the identification of $\dot\theta$ with baryonic chemical potential, $\dot\theta = \mu_B$, is questionable, as we argue below. 

\subsection{General consideration} \label{ss-general}

In the canonical approach the Hamiltonian density, ${\cal H}$, is derived from the Lagrangian density, ${\cal L}$, in the following way. 
The Lagrangian density is supposed to depend upon some field variables, $\phi _a $, and their first derivatives, 
$\partial _{\mu} \phi _a$.
First, we need to define the canonical momentum conjugated to the "coordinate" $\phi _a$:
\be
\pi_{a} = \frac{\partial {\cal L}}{\partial \dot \phi_a}\, .
\label{pi-a}
\ee  
The Hamiltonian density is expressed through the canonical momenta and coordinates as 
\be 
{\cal H} = \sum_a \pi_{a} \dot \phi _a - {\cal L}\, ,
\label{can-H}
\ee
where the time derivatives, $\dot \phi _a$, should be written in terms of the canonical momenta, $\pi _a$.  

The Hamilton equations of motion:
\be
\dot  \phi = \frac{\partial {\cal H}}{\partial \pi}\,\,\,{\rm and} \,\,\, \dot  \pi = - \frac{\partial {\cal H}}{\partial \phi}
\label{ham-eqs}
\ee
are normally equivalent to the Lagrange equations obtained by the least action principle from the Lagrangian.

For example for a real scalar field with the Lagrangian
\be
{\cal L} (\chi) =  (\partial \chi)^2/2  - m^2_\chi  \chi^2 /2
\label{L-of-chi}
\ee
the canonical momentum is $\pi_\chi = \dot\chi$ and the Hamiltonian density is:
\be
{\cal H} (\chi) = (1/2) \left[ \pi_ \chi^2 + (\nabla \chi)^2 + m^2_\chi \chi^2 \right]  ,
\label{H-of-phi}
\ee
while for a complex scalar field with
\be
{\cal L} (\phi) = | \partial \phi |^2  - m^2_\phi | \phi |^2
\label{L-of-phi}
\ee
the canonical momenta are $\pi_\phi = \dot \phi^*$ and $\pi_{\phi^*} = \dot \phi $ and the Hamiltonian density is:
\be
{\cal H} (\phi) = \pi_\phi \pi_{\phi^*}+ |\nabla \phi |^2 + m_\phi^2 | \phi |^2\, .
\label{H-of-phi}
\ee
The corresponding Hamilton equations lead, as expected,  to the usual Klein-Gordon equations for  $\phi$ or $\chi$.

For the Dirac field with
\be
{\cal L}(\psi)  = \bar \psi \left( i \partial \!\!\! /  - m_\psi \right) \psi
\label{L-of-psi}
\ee
the canonical momenta are  $\pi_\psi = i \psi^\dagger$ and $\pi_{\psi^\dagger} = 0$, so we arrive to the well known expression:
\be
{\cal H} (\psi) = \psi^\dagger \left( i \gamma_4 \gamma_k \partial_k + \gamma_4 m \right) \psi .
\label{H-of-psi}
\ee
Let us make now the same exercise but with the symmetric Lagrangian, which differs from the canonical one by a total derivative:
\be
{\cal L}_{sym} (\psi) = \left[ \bar \psi \left( i \partial \!\!\! /  - 2m_\psi \right) \psi - i (\partial_\mu \bar \psi) \gamma_\mu \psi  \right]/2 .
\label{L-of-psi-s}
\ee
The corresponding canonical momenta are: $\pi_\psi = i \psi^\dagger/2$ and  $\pi_{\psi^\dagger} = - i \psi/2$ and the Hamiltonian 
density is
\be 
{\cal H}_{sym} (\psi) = m_\psi \psi^\dagger \gamma_4 \psi +  \frac{i}{2} \left( \psi^\dagger  \gamma_4 \gamma_k \partial_k \psi -
\partial_k \psi^\dagger \gamma_4 \gamma_k  \psi \right),
\label{H-of-psi-s}
\ee
which differs from the usual expression (\ref{H-of-psi}) by the space divergence, 
$ (i/2) \partial_k ( \psi^\dagger \gamma_4 \gamma_k \psi)$.  The total Hamiltonian, defined as 
\be
H = \int d^3 x\, {\cal H}, 
\label{H-tot}
\ee 
remains the same in both cases, (\ref{H-of-psi}) and (\ref{H-of-psi-s}),  if the fields vanish at spatial infinity.
Below the field $\theta$ depending only on time  is considered, but one can assume that it weakly depends upon
the space coordinates and vanishes at infinity. The local dynamics in this case remains undisturbed.

\subsection{The case of SSB} \label{ss-SSB}

Let us consider now a model with the coupling 
\be
{\cal L}_{SB}(\Theta) =  (\partial_\mu  \Theta)  J^{\mu}_B , 
\label{L-SB-1}
\ee
where $\Theta$ is some scalar field and  $J^\mu_B$ is 
a vector baryonic current. It has the form:
\be
J^\mu_B = B\, \bar \psi \gamma^\mu \psi ,
\label{j-mu-Q}
\ee
where $\psi$ is some fermionic baryon (e.g. quark) and $B$ is its baryonic number. Such interaction is postulated in spontaneous 
baryogenesis scenarios~\cite{spont-BG-1,spont-BG-2,spont-BG-3,BG-rev}
or in gravitational baryogenesis~\cite{gravBG-1,gravBG-papers}.  In the former case $\Theta = \theta $ is a (preudo)goldstone field, while in the 
latter $\Theta = R/m_R^2$ with $R$ being the curvature scalar and $m_R$ is a constant parameter with dimension of mass.

In what follows we confine ourselves to consideration of the Goldstone field $\theta$ and distinguish between
the following two possibilities: 
\begin{enumerate}[A.]
\item
 $\theta$ is a dynamical field with the free Lagrangian of the form given by Eq.~(\ref{L-of-chi}) where $\chi = f\, \theta$. This is exactly the situation which is realized 
 in the case of spontaneous symmetry breaking.
 \item
 $\theta$ is an external "fixed" field. The term "fixed" is used here in the sense that the dependence of $\theta$ on
coordinates is fixed by some dynamics which does not enter into the Lagrangians under scrutiny. This is the case which is studied 
both in the spontaneous baryogenesis and in the gravitational baryogenesis. It is considered in the next subsection. 
\end{enumerate}

In the canonical case A the Hamiltonian density is calculated in accordance with the specified above rules. 
Correspondingly, for the Lagrangian (\ref{L-of-theta-1}) we obtain: 
  \be
{\cal H}_1 (\theta)
= {f^2\over 2} \left(\dot \theta^2 + (\nabla \theta)^2 \right) +  
Q_1^\dagger \gamma_4  (i \gamma_k \partial_k + m_Q) Q_1 +  
 L ^\dagger \gamma_4 (
i \gamma_k \partial_k + m_L) L - \nonumber \\
 \left(\frac{e^{ - i \theta }}{ m_X^2} \, ( Q_1^\dagger \gamma_4 \gamma_{\mu} L )( Q_1^\dagger \gamma_4  \gamma_{\mu} Q_1^c) +
  h.c.\right) +
 U(\theta)\, ,
 \label{H-of-theta-1}
 \ee
where the $\theta $-conjugated canonical momentum is $\pi _{1 \theta} = f^2 \dot \theta $. 

Analogously for the Lagrangian (\ref{L-of-theta-2}) the Hamiltonian density is:
  \be
{\cal H}_2 (\theta)
= {f^2\over 2} \left(\dot \theta^2 + (\nabla \theta)^2 \right) +  
Q_2^\dagger \gamma_4  (i \gamma_k \partial_k + m_Q) Q_2 +  
 L ^\dagger \gamma_4 (
i \gamma_k \partial_k + m_L) L - \nonumber \\
 \left(\frac{1}{ m_X^2} \, ( Q_2^\dagger \gamma_4 \gamma_{\mu} L )( Q_2^\dagger \gamma_4  \gamma_{\mu} Q_2^c) +
  h.c.\right) + U(\theta) - (1/3) (\partial_k \theta) (Q_2^\dagger \gamma_4  \gamma_k Q_2) \, ,
 \label{H-of-theta-2}
 \ee
where the canonical momentum is $\pi _{2 \theta} = f^2 \dot \theta + n_B$. Correspondingly,  
$\dot \theta $ should be expressed through the canonical momentum $\pi _{2 \theta}$ according to
\be  
\dot \theta = (\pi _{2 \theta} - n_B)/f^2 .
\label{dot-theta-pi}
\ee
Taking into account that $Q_1= e^{- i\theta/3}Q_2$ we can check that the Hamiltonians (\ref{H-of-theta-1}) and 
(\ref{H-of-theta-2}) interchange under this transformation. Thus we see that the calculation of Hamiltonians according
to the specified rules is self-consistent.

Note, that both Hamiltonians, as they are presented in eqs.~(\ref{H-of-theta-1}) and (\ref{H-of-theta-2}),  
do not contain "chemical potential", $\dot\theta$,  in the form $\dot \theta n_B$ and in this sense 
contradict  the presumption (\ref{H-SB}). However, the case is somewhat more tricky. 
Written in terms of the canonical momentum the corresponding part of the Hamiltonian (\ref{H-of-theta-2}) (the first term) has the form 
$\delta {\cal H}_2 (\theta) = (\pi_{2\theta} - n_B)^2/(2f^2) $. 
In spatially independent case and in absence of $U(\theta)$
the Hamiltonian equation of motion for ${\cal H}_2$ 
has the form $\dot \pi_{2\theta}  = 0$, so its solution is $\pi_{2 \theta} = const$. 
Evidently this equation is equivalent to the Lagrange
equation of motion for $\theta$-field (\ref{d2-theta-2}) (where the cosmological expansion is neglected).

The presence of $(-\pi_{2\theta} n_B /f^2)$ - term in the Hamiltonian (\ref{H-of-theta-2}) implies 
that   $\pi_{2\theta} /f^2$ can be understood as the baryonic chemical potential, $\mu_B$.
Since it is usually assumed that initially $n_B (t_{in})= 0$, then $\pi_{2\theta} = f^2 \dot\theta (t_{in})$ and thus
$\mu_B = \dot\theta (t_{in})$, but not $\mu_B=\dot \theta (t)$ taken at the running $t$ for which thermal equilibrium is established.

\subsection{External field $\theta$ \label{ss-external} }

The assertion  (\ref{H-SB}) might be in principle valid, if $\theta $ was an external "fixed" field with the
dynamics determined "by hand", as it is noted 
in  subsection \ref{ss-SSB}. In this case expression~(\ref{H-SB}) could be formally true but, as we show here, 
such a theory possibly has some internal inconsistencies. 

Let us study previously considered theories with Lagrangians  (\ref{L-of-theta-1}) and (\ref{L-of-theta-2}),
where the kinetic and potential terms 
for $\theta $ are omitted. We have two options for construction of Hamiltonians: either to proceed along the usual lines specified above or to assume 
the validity of the prescription ${\cal H}_{int} = - {\cal L}_{int}$ for the interaction parts of Lagrangians. There is an unambiguous procedure for 
Lagrangian (\ref{L-of-theta-1}), since its interaction part does not contain derivatives. It is not so for Lagrangian (\ref{L-of-theta-2}), because of 
the term $(\partial_{\mu} \theta) J^{\mu}_B$ for which the conjecture ${\cal H}_{int} = - {\cal L}_{int}$ is not true. As we have seen in 
subsection \ref{ss-SSB}, the standard approach leads to the Hamiltonian (\ref{H-of-theta-2})
which does not contain the term $\dot \theta (t) n_B$. To arrive to the mechanism of spontaneous baryogenesis described in the 
literature we need to postulate 
${\cal H}_{int} = - {\cal L}_{int}$ independently on the presence of the field derivatives. If this postulate
was true, the Lagrangian (\ref{L-of-theta-2}) would lead to Hamiltonian containing the necessary term $\dot \theta n_B$. On the other hand, if we 
apply the standard procedure to calculate the Hamiltonian from the Lagrangian without the kinetic term, we find $\pi_\theta = n_B$ and arrive to the
striking result:
\be  
{\cal H} (\theta) =  \pi_\theta \dot\theta - {\cal L}  = 0,
\label{H-0}
\ee
which clearly demonstrates an inconsistency of a theory without the kinetic term.  

Additional problems appear if we consider the theory with the Lagrangian 
\be
{\cal L}_{SB}^{(1)} =   -\theta \partial _\mu J^\mu_B \, ,
\label{L-SB-1}
\ee
which differs from the original ${\cal L}_{SB}$ (\ref{L-SB}) by the total divergence and thus  
leads to the same Lagrangian equations of motion, so these Lagrangians are physically equivalent. However, 
it may be not so for the Hamiltonian densities.
The Lagrangian (\ref{L-SB-1}) does not contain the time derivative of the theta-field but contains time derivatives of the dynamical
fermionic fields. So the Hamiltonian obtained from  ${\cal L}_{SB}^{(1)} $ through the specified above
standard rules, applied to fermions, has the form: 
\be 
{\cal H}_{SB}^{(1)}  = ( \partial _k \theta) J_k - \partial _k (\theta J_k)  \rar  ( \partial _k \theta) J_k\, ,
\label{H-SB-1}
\ee
where at the last step we omitted the spatial divergence. Evidently the Hamiltonian ${\cal H}_{SB}^{(1)} $ differs 
from ${\cal H}_{SB}$ (\ref{H-SB}), 
though they are obtained from the equivalent Lagrangians. It means that the Hamiltonian equations of motion 
corresponding to ${\cal H}_{SB}$ and ${\cal H}_{SB}^{(1)}$ would be different.  It can be checked that the
equations derived from the Hamiltonian (\ref{H-SB-1}) disagrees with the Lagrangian ones.   
However, this  is not the problem inherent to SBG but to the problem with the determination of the Hamiltonian density 
of the fermionic fields, related to the degeneracy between the coordinate $\psi$ and the canonical momentum 
$\psi^\dagger$, see sec.~\ref{ss-general}.
These problems will be considered elsewhere, while in this work we concentrate on the kinetics of the 
standard scenario of SBG, which in many cases leads essentially to the usual results presented in the literature.
However, this is not always so.

\section{Kinetic equation for time independent amplitude \label{s-kin-eq-canon}}

\subsection{Kinetic equilibrium \label{ss-kin-equil}}

The study of kinetics of fermions in the cosmological background is grossly simplified if the particles
are in equilibrium with  respect to elastic scattering, to their possible annihilation e.g. into photons, and
to other  baryo-conserving interactions.
The equilibrium with respect to elastic scattering implies the following form of the phase space distribution 
functions:
\be
f_{eq} =\left[ 1 +  \exp (E/T -\xi) \right]^{-1},
\label{f-eq-ferm}
\ee
where the dimensionless chemical potential $\xi = \mu/T$ has equal magnitude but opposite signs for particles and antiparticles.
The baryonic number density for small $\xi$ is usually given by the expression
\be 
n_B = g_S  B_Q \xi_B T^3 /6 
\label{nB-of-xi}
\ee
(compare to eq. (\ref{n-B-of-theta})). 
Here $\xi_B$ is the baryonic chemical potential. 
This  equation which expresses baryonic number density through chemical potential 
is true only for the normal relation between the energy and three-momentum, $E=\sqrt{ p^2 + m^2}$, 
with equal masses of particles and antiparticles.

Vanishing baryon asymmetry
implies $\xi_B = 0$, as is usually the case. If the baryonic number of quarks is conserved, 
$n_B$ remains constant in the comoving volume and it means in turn that $\xi = const$ for massless
particles. If  $n_B=0$  initially, then $\xi_B$ remains identically zero. 
If baryonic number is not conserved, then 
as we see below from the kinetic equation, equilibrium with respect to B-nonconserving processes leads to $  \xi_B = c\,\dot\theta/ T$,
as is envisaged by SBG. The constant $c$ depends upon the concrete type of reaction.
Complete thermal equilibrium
in the standard theory demands $n_B \rar 0$, but a deviation from thermal equilibrium of B-nonconserving interaction leads to generation 
of non-zero $\xi_B$ and correspondingly to non-zero~$n_B$. 

The situation changes, if quarks and antiquarks satisfy the equation of motion (\ref{dirac-2}), for which the
following dispersion relation is valid
\be 
E  = \sqrt{p^2 +m^2} \mp \dot\theta/3,
\label{E-split}
\ee
where the signs $\mp$ refer to particles or antiparticles respectively. So the energies of quarks and antiquarks
with the same three-momentum are different. This is similar to mass difference which may be induced by CPT violation.
It is noteworthy that the above dispersion relation is derived under assumption of constant or 
slow varying $\dot\theta$. Otherwise the Fourier transformed Dirac equation cannot be reduced to the algebraic one and
the particle energy is not well defined.

The baryon number density corresponding to the dispersion relation (\ref{E-split}) is given by the expression
\be
n_B \equiv g_S  B_Q  \int \frac{d^3 p }{(2\pi)^3} \left[ f (p) - \bar f (p) \right] = 
\frac{ g_S  B_Q}{6} \left(\xi_B + \frac{\dot\theta}{3T}\right)\,T^3 .
\label{n-B-E-split}
\ee
where $\bar f$ is the distribution function of antiparticles. If the baryon number is conserved and is zero initially, 
the condition $ \xi_B  + \dot\theta/(3T) =0$
would be always fulfilled. If B is not conserved, then the equilibrium with respect to B-nonconserving processes demands
$\xi_B = 0$, as it follows from kinetic equation presented below. 
So evidently $\xi_B\neq \dot \theta$ but nevertheless the baryon asymmetry is proportional to~$\dot\theta$ as follows from eq.~(\ref{n-B-E-split}).

\subsection{Relation between $n_B(t)$ and $\theta(t)$ in the pure goldstone  case \label{ss-nB-theta} }

Equation of motion for theta-field in cosmological background (\ref{d2-theta-FRW}) with $U(\theta)=0$
can be easily integrated expressing
baryon asymmetry, $n_B$, through $\dot \theta$. In the case when the relation (\ref{nB-of-xi}) is fulfilled, we obtain: 
\be
f^2  \left[ \frac{\dot\theta (t)}{T^3(t) }- \frac{\dot\theta (t_{in})}{T_{in}^3} \right] = - \frac{ g_S B_Q}{6} \left[  \xi_B (t) - \xi_B (t_{in}) \right] ,
\label{dot-theta-of-xi}
\ee
assuming  that the temperature drops according to the law $\dot T = - HT$.

The initial  value of the baryon asymmetry is usually taken to be zero,
so according to eq.~(\ref{nB-of-xi}) we should also take 
 $\xi_B (t_{in}) = 0$. Let us remind that eq.~(\ref{nB-of-xi}) is valid for the case of normal dispersion relation, 
 $E=p$ (in massless case), both for quarks and antiquarks.   
    
In the theory with the Lagrangian (\ref{L-of-theta-2}) and with the Dirac equation (\ref{dirac-2}) the dispersion relation changes to 
(\ref {E-split}) and the relation between $n_B$ and $\xi_B$ becomes (\ref{n-B-E-split}). Now eq.~(\ref{d2-theta-FRW}) is integrated as:      
\be
f^2  \left[ \frac{\dot\theta (t)}{T^3(t) }- \frac{\dot\theta (t_{in})}{T_{in}^3} \right] = - \frac{ g_S B_Q}{6} \left[  \xi_B (t) - \xi_B (t_{in}) 
+ \frac{\dot \theta (t)}{3T} - \frac{\dot \theta (t_{in})}{3T_{in}}
\right] .
\label{dot-theta-of-xi-2}
\ee
If initially $n_B = 0$, then $\xi_B (t_{in}) = -\dot\theta_{in}/(3T_{in})$. 

In the pseudogoldstone case, when  $U(\theta) \neq 0$, equations of motion (\ref{d2-theta-2}) or (\ref{d2-theta-FRW}) 
cannot be so easily integrated, but in thermal equilibrium the system 
of equations containing $\theta (t)$ and $\xi_B(t)$ can be reduced to ordinary differential equations which are easily solved numerically.
Out of equilibrium one has to solve much more complicated system of the ordinary differential equation of motion for $\theta(t)$
and the integro-differential kinetic equation. It is discussed below in sec.~\ref{s-kin-eq-canon}.

\subsection{Kinetic equation in (quasi)stationary background \label{ss-kin-eq-stat}}

The probability of any reaction between particles in quantum field theory is determined by the amplitude 
of transition from an initial state $| in \rangle$ to a final state $| fin \rangle$. 
In the lowest order of perturbation theory the transition amplitude  is
given by the integral of the matrix element of the Lagrangian density between these states, integrated over 4-dimensional space $d^4 x$.
Typically the quantum field operators are expanded in terms of creation-annihilation operators with the plane wave coefficients as:
\be
\psi (t,{\bf x}) =
	\int \frac{d^3q}{2E \, {(2\pi)^3}} 
	\left[ a ({\bf q})  e^{-i q x} + b^\dag ({\bf q})  e^{iqx}
	\right].
\ee
where $a$, $b$, and their conjugate are the 
annihilation (creation) operators for spinor particles and antiparticles and $q x = E t - {\bf q x} $. 

If the amplitude of the process is time-independent, then
the integration over $dt d^3 x$ of the product of the exponents of $i q x$ in infinite integration limits
leads to the energy-momentum conservation factors:
\be
\int dt d^3 x \,e^{-i ( E_{in} - E_{fin} ) t + i (\bf P_{in} - \bf P_{fin} ) \bf x } = (2\pi)^4 \delta (E_{in} - E_{fin})\, \delta((\bf P_{in} - \bf P_{fin} ),
\label{int-over-d4x-1}
\ee
where $E_{in} $, $E_{fin}$,    $\bf P_{in}$, and $\bf P_{in}$ are the total energies and 3-momenta of the initial and final states respectively.
The amplitude squared contains delta-function of zero which is interpreted as the total time duration, $t_{max}$, of the process and as the
total space volume, $V$. The probability of the process given by the collision integral is normalized per unit time and volume, so it must be 
divided by $V$ and $t_{max}$.

The temporal evolution of the distribution function 
of i-th type particle, $f_i (t,p)$, in an arbitrary process
${ i+Y \lrar Z}$ in the FRW background, is governed by the equation:
\be
\frac{df_i}{dt} = (\partial_t - H\,p_i \partial_{p_i}) f_i = I_i^{coll}, 
\label{kin-eq-gnrl}
\ee
with the collision integral equal to:
\be 
&&I^{coll}_i={(2\pi)^4 \over 2E_i}  \sum_{Z,Y}  \int  \,d\nu_Z  \,d\nu_Y
\delta^4 
(p_i +p_Y -p_Z)
\nonumber\\
&& \left[ |A(Z\rightarrow  i+Y)|^2
\prod_Z f \prod_{i+Y} (1\pm f) - 
 |A(i+Y\rightarrow  Z)|^2
 f_i  \prod_Y  f\prod_Z  (1\pm  f)\right],
\label{I-coll}
\ee
where $ A( a \rar b)$ is the amplitude of the transition from state $a$ to state $b$,
${ Y}$ and ${ Z}$ are  arbitrary, generally  multi-particle  states,
$\left( \prod_Y f \right)$ is  the  product  of  the phase space densities  of  particles forming the state $Y$, and
\be
d\nu_Y = \prod_Y {\overline {dp}} \equiv \prod_Y {d^3p\over (2\pi )^3 2E}.
\label{dnuy}
\ee
The signs '+' or '$-$' in ${ \prod (1\pm f)}$ are chosen for  bosons  and fermions respectively. We neglect the effects of space-time
curvature in the collision integral, which is generally a good approximation.

We are interested in the evolution of the baryon number density, which is the time component of the baryonic current
$J^\mu$: $n_B \equiv J^4$. Due to the quark-lepton transitions the current is non-conserved and its divergence is
given by eq.~(\ref{dmu-Jmu}). The similar expression is evidently true in terms of $Q_2$ but without the factor $\exp (- i \theta)$.
Let us first consider  the latter case, when
the interaction described by the Lagrangian (\ref{L-of-theta-2}), which contains the product of three "quark" and one "lepton"
operators, and take as an example the process  $q_1 + q_2 \lrar \bar q + l $. 

Since the interaction in this representation does not depend on time, the energy is conserved and the collision integral has the 
usual form with conserved four-momentum. Quarks are supposed to be in kinetic equilibrium but probably not in equilibrium with 
respect to  B-nonconserving interactions, so their distribution functions have the  form:
\be
f_q = \exp \left( -\frac{E}{T} + \xi_B \right) \,\,\, {\rm and} \,\,\, f_{\bar q} = \exp \left( -\frac{E}{T} - \xi_B \right) 
\label{f-B-equil}.
\ee
Here  and in what follows the Boltzmann statistics is used. According to ref.~\cite{ad-kk}, 
Fermi corrections are typically at the 10\% level. Since the dispersion relation for quarks and antiquarks ~(\ref{E-split})
depends upon $\dot\theta$, the baryon asymmetry in this case is given by eq.~(\ref{n-B-E-split}) and  the kinetic equation takes the form:
\be
\frac{g_S  B_Q}{6}\, \frac{ d}{dt} \left( \xi_B + \frac{\dot\theta}{3T}\right) = - c_1\Gamma \xi_B,
\label{dot-n-B-xi}
\ee
where $c_1$ is a numerical factor of order unity and $\Gamma$ is the rate of baryo-nonconserving reactions.
If the amplitude of this reaction has the form  determined by the Lagrangian~(\ref{L-of-theta-2}), then $\Gamma \sim T^5/m_X^4$.

For constant or slow varying temperature the equilibrium solution to this equation is $\xi_B = 0$ and the baryon number
density (\ref{n-B-E-split}) is proportional to $\dot \theta$,  $n_B =(g_S B_Q/18)\, \dot\theta T^2 $, with $\dot\theta$ evolving 
according to eq. (\ref{dot-theta-of-xi-2}) as:
\be
\dot \theta = \frac{ f^2 }{f^2 + g_S B_Q T^2 /18}\, \left(\frac{T}{T_{in}}\right)^3\,\dot\theta( t_{in}) .
\label{dot-theta-anom}
\ee
We see that the equilibrium value of $n_B$ drops down with decreasing temperature as $T^5$. However at small temperatures 
baryon non-conserving processes switch-off  and $n_B$ tends to a constant value in comoving volume.

Let us check now what happens if the dependence on $\theta$ is moved from the quark dispersion relation to the B-nonconserving
interaction term (\ref{d2-theta-1}).
The collision integral (\ref{I-coll}) contains delta-functions imposing conservation of energy and momentum if there is no external field 
which depends upon coordinates.  
In our case, when 
quarks "live" in the $\theta (t)$-field, the collision integral should be modified in the following way. 
 We have now an additional factor under integral (\ref{int-over-d4x-1}), namely, $\exp [ \pm i\theta (t)]$. 
 In general case this integral cannot be 
taken analytically, but if we can approximate $\theta (t) $ as $\theta(t) \approx \dot \theta t$ with a constant or slowly varying 
$\dot \theta$, the integral is simply taken giving e.g. for the process of two quark transformation into antiquark and lepton,
$q_1+q_2 \lrar \bar q + l$, the energy balance condition imposed by
$\delta (E_{q_1} + E_{q_2} - E_{\bar q} - E_l - \dot\theta)$.
In other words the energy is non-conserved due to the action of the external field $\theta (t)$. The approximation of linear evolution of 
$\theta $ with time can be valid if the reactions are fast in comparison with the rate of the $\theta$-evolution. 

Note in passing that with a non-zero $\theta (t)$ the current non-conservation (\ref{dmu-Jmu}) in principle may induce baryogenensis
because it breaks not only baryonic number conservation, but also CP, due to complexity of the coefficients. However, in this
particular model no baryon asymmetry would be generated. The model is quite similar to the model of the baryon asymmetry generation
in heavy particle decays, such as e.g. GUT baryogenesis. However, as it is argued e.g. in Refs.~\cite{ad-yab-cpt,Kolb}, for the generation
of the asymmetry at least three different channels of baryo-nonconserved reactions are necessary. Thus one would need to add
some extra fields into the model to activate  this mechanism.

Returning to our case we can see that the collision integral
taken over the  three-momentum of the particle under scrutiny
(i.e. particle $i$ in eq.~(\ref{I-coll}) )
 e.g. for process the $q_1+q_2 \rar  l + \bar q $ turns into:
\be
\dot n_B +3H n_B \sim \int d\tau_{l \bar q} d\tau_{q_1 q_2} |A|^2 \delta (E_{q_1} + E_{q_2} - E_l - E_{\bar q} - \dot \theta) 
\delta( {\bf P}_{in} - {\bf P}_{fin} ) e^{- E_{in}/T} 
\left( e^{\xi_L - \xi_B + \dot \theta/T} - e^{2 \xi_B} \right),
\label{dot-nB-1}
\ee
where $d\tau_{l,\bar q} = d^3 p_l d^3 p_{\bar q} /[ 4 E_l E_{\bar q} (2\pi)^6]$. 
We assumed here that all participating particles are in kinetic equilibrium, i.e. their distribution functions have the 
form (\ref{f-B-equil}).
In expression (\ref{dot-nB-1}) $\xi_B$ and $\xi_L$ denote baryonic and leptonic chemical potentials respectively and
the effects of quantum statistics are neglected but only for brevity of notations. The assumption of kinetic equilibrium is well justified because it is enforced by the very efficient elastic scattering.
Another implicit assumption is the usual equilibrium relation between chemical potentials of particles and antiparticles,
$\bar \mu = -\mu$, imposed e.g. by the fast annihilation of quark-antiquark or lepton-antilepton pairs into two and three photons.
Anyhow the assumption of kinetic equilibrium is one of the cornerstones of the spontaneous baryogenesis.

The conservation of $(B+L)$ implies the following relation: $\xi_L = - \xi_B/3$. Keeping this in mind, we find
\be
\dot n_B + 3H n_B \approx  -\left(1 - e^{ \dot \theta/T - 3\xi_B +\xi_L} \right) I  \approx   \left( \frac{\dot \theta }{T} - 
\frac{10}{3}\,\xi_B \right) I,
\label{dot-nB-2}
\ee
where we assumed that $\xi_B$ and $\dot \theta/T$ are small.
In relativistic  plasma with temperature $T$ the factor $I$, coming from the collision integral, 
can be estimated as $I=T^8/m^4$, where $m$ is a numerical constant with dimension of mass. 
It differs from $m_X$, introduced in eq.~(\ref{L-of-theta-1}), by a numerical coefficient.

For a large factor $I$  we expect the equilibrium solution 
\be
\xi_B = \frac{3}{10} \frac{\dot\theta}{T},
\label{xi-B-L} 
\ee
so $\dot\theta$ up to the numerical
factor seems to  be the baryonic chemical potential, as expected in the usually assumed SBG scenario. 
The value of the coefficient $c=3/10$ in eq.~(\ref{xi-B-L}) may be different for other types of B-nonconcerving reactions,  
e.g. for the reaction $3q \lrar 3 \bar q$ one can find that $c=1/6$. 
Let us remind that for the dispersion relation (\ref{E-split}) 
 the baryonic chemical potential is not  proportional to $\dot\theta (t)$, but is equal to zero, see eq.~(\ref{dot-n-B-xi}) 
and comments below. 

\section{Out-of-equilibrium generation of baryon asymmetry in purely Goldstone case \label{s-evol-theta-G}}

As we have seen in the previous section the equilibrium value of the baryon asymmetry in comoving volume drops down as
$T^2$. So for an effective generation of the asymmetry the B-nonconserving reactions must drop out  of equilibrium at sufficiently high 
temperatures. Below we estimate the asymptotic value of the baryon asymmetry.  

Let us first study the case when the cosmological expansion is very slow and the temperature can be considered as constant or, 
better to say, adiabatically decreasing.
The proper equations in this limit can be solved analytically and it allows a better insight into the problem. With constant $T$ the equilibrium 
would be ultimately reached if time is sufficiently large
and asymptotically the baryonic chemical potential is indeed proportional to $\dot\theta (t)$, but one should remember that this is true
in the case when $\theta (t)$ enters the interaction term but not the quark dispersion relation.
Similar situation is realized in cosmology with decreasing  temperature of the cosmic plasma but it is interesting
that the magnitude of the resulting baryon asymmetry is a non-monotonic function of the strength of B-violation. With very strong
and very weak interaction the  asymmetry goes to zero and 
the best conditions for baryogenesis are realized in the intermediate case.

Using eqs.~(\ref{nB-of-xi}), (\ref{dot-theta-of-xi}), and (\ref{dot-nB-2}) we find
\be
\dot \xi_B = \gamma \left[ \frac{\dot \theta_{in} }{T} - \xi_B \left(\frac{10}{3} + \frac{C_B T^2}{f^2} \right)\right],
\label{dot-xi-5}
\ee
which is solved as
\be
\xi_B(t) = \frac{\dot\theta_{in}}{T \kappa} \left[ 1 - e^{- \kappa \gamma ( t - t_{in} )} \right] ,
\label{xi-B-of-t}
\ee
where $C_B = g_S B_Q/18$, 
 $\gamma = T^5/(C_B m^4)$, $\kappa =  10/3 +C_B T^2/f^2$,
$t_{in}$ is the initial value of time, at which $\xi_B (t_{in})= 0$, and $\dot\theta_{in} = \dot\theta (t_{in}) $.

The time derivative of the Goldstone field evolves as
\be
\dot\theta (t) = \dot\theta_{in} \left[ 1 - \frac{C_B T^2}{f^2 \kappa} \left( 1 - e^{-\kappa\gamma (t - t_{in} )} \right) \right] .
\label{dot-theta-of-t}
\ee
So $\dot \theta (t)$ drops down  asymptotically at large time with respect to its initial value, and the baryonic chemical potential exponentially  
tends to $\xi_B \rar  \dot\theta_{in} /(\kappa T )$,
as it is expected in SBG scenario. 

As follows from eq. (\ref{dot-theta-of-t}), $\dot\theta$ tends to a constant value at large $t$, however at the beginning the second
time derivative $ \ddot\theta $ may be non-negligible:
\be
\ddot\theta = -\frac{\dot\theta_{in} C_B T^2 \gamma}{f^2}  e^{-\kappa\gamma (t - t_{in}) } .
\label{ddot-theta-in}
\ee
The variation of $\dot\theta $ with time is considered in sec. \ref{ss-Taylor}.

Let us turn now to more realistic cosmology when the temperature  drops down according to 
\be 
\dot T = - H T
\label{dot-T}
\ee
with the Hubble parameter equal to
\be
H = \left( \frac{8 \pi^3 g_*}{90}\right)^{1/2} \frac{T^2}{m_{Pl}} \equiv G_* \frac{T^2}{m_{Pl}},
\label{H-of-T}
\ee
where $m_{Pl} = 1.2 \times 10^{19} $ GeV is the Planck mass and $g_*$ is the number of species in the primeval relativistic plasma. 
In the interesting temperature range $g_* \sim 100$. 

Now $\dot\theta (t)$ is expressed through $\xi_B(t) $ according to  eq. (\ref{dot-theta-of-xi}) and instead of eq.~(\ref{dot-xi-5})
we obtain:
\be
\dot \xi_B = \gamma \left( \frac{\dot \theta_{in}  T^2}{T_{in}^3} - \kappa \xi_B  \right)
\label{dot-xi-B5}
\ee
This equation can be more conveniently solved if we change time variable as $dt = - dT/(HT)$ and introduce dimensionless inverse temperature
according to $\eta = T_{in}/T$. So the baryonic chemical potential evolves as a function of $\eta = T_{in}/T$ as:
\be
\xi_B (\eta) = K \int_1^\eta \frac{d\eta'}{(\eta')^6}\,\exp\left[ - N\int_{\eta'}^\eta \frac{d\eta''}{(\eta'')^4}
\left( \frac{10}{3} + \frac{C_B  T_{in}^2}{f^2 \eta''^2}\right) \right],
\label{xi-B-of-eta}
\ee
where $K = \dot\theta_{in} m_{Pl} T_{in}^2/({C_B\,m^4G_*})$, $N=m_{Pl} T_{in}^3/({C_B\,m^4G_*})$. 
If $K \gg 1$, which corresponds to the equilibrium case, 
the integral can be evaluated up to 
the terms of the order of $1/K$ and we find:
\be
\xi_B (\eta) = \frac{(\dot\theta_{in} /T_{in})}{ (10\eta^2 /3) + (C_B T_{in}^2) / f^2 }\,.
\label{xi-of-eta-asymp}
\ee
This result coincides, as expected, with the equilibrium solution of eq.~(\ref{dot-xi-B5}): $ \xi_B= \dot \theta_{in}\,T^2/(T^3_{in} \kappa )$.
Note that in equilibrium both $\xi_B $ and $\dot\theta /T$ fall down
as $T^2$ with decreasing temperatures.

It is instructive to consider a different model of baryonic number non-conservation through quark-antiquark transformation
$ 2Q \lrar 2\bar Q $. For realistic quarks such process is forbidden, but the process $ 3q \lrar 3\bar q $ is allowed in e.g. $SO(10)$
model of grand unification. However, we consider the first one just for simplicity. The kinetic equation (\ref{dot-nB-2})
in this case is transformed into:
\be 
\dot n_B = \left( \frac{\dot\theta}{T} - 4 \xi_B \right) \,\frac{T^8}{m^4},
\label{dot-nB-4}
\ee
so in equilibrium with respect to the process $2Q \lrar 2\bar Q$ the baryonic chemical potential tends to
$\xi_B \rar \dot\theta / (4T )$.

\begin{figure}[htbp] 
\centering
 \includegraphics[width=.40\textwidth]{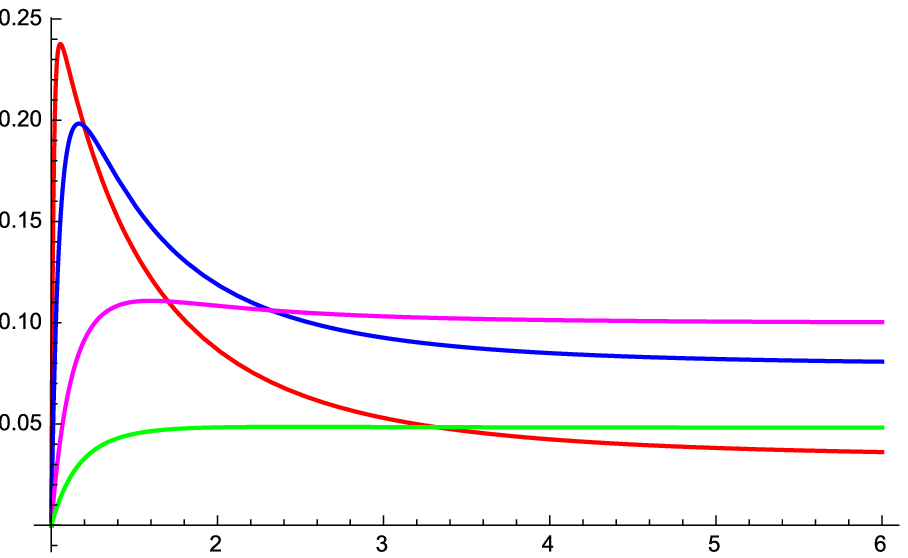} \hspace{.2cm}
\includegraphics[width=.40\textwidth]{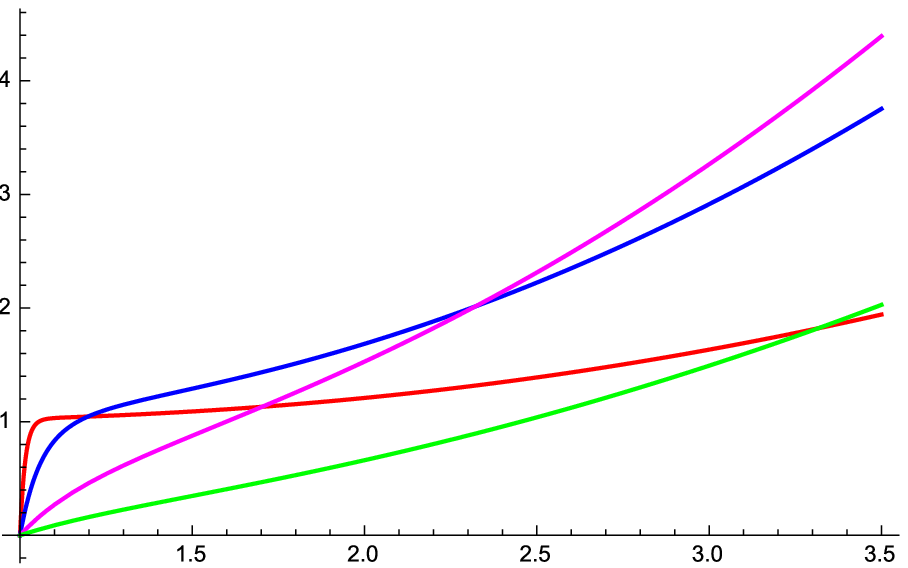}
\caption{
{\it Left}:  
Evolution of $\xi_B (\eta)$ according to eq.~(\ref{xi-B-of-eta}) 
where $ C_B T_ {in}^2 /(5 f^2) =   0.1$ for  K = N =  20 (red),  5 (blue),  1 (magenta), and 0.3 (green).  
{\it Right}: Ratio of $\xi_B (\eta)$ to its equilibrium value (\ref{xi-of-eta-asymp})  for the same values of the parameters.}
\label{fig-1}       
\end{figure}

Now we will see what happens out of equilibrium. To this end we numerically take the integral  in eq. (\ref{xi-B-of-eta}) for different 
values of $K$ and $C_B T_{in}^2/f^2$.
The results for $\xi_B(\eta) $ and the ratio of $\xi_B$ to the equilibrium value  $(3/10) \dot\theta /T$ as functions of 
$\eta = T_{in}/T$ are presented in Fig.~\ref{fig-1}, in  left and right panels respectively.
As is seen from the left panel, the baryon asymmetry is a non-monotonic function of the rate of the baryo-nonconserving processes. 
For a large rate (large K and N) baryon asymmetry is quickly generated and reaches high value, but it drops down as the 
equilibrium one, $\sim 1/\eta^2$, till lower temperatures. As a result the final baryon asymmetry is smaller for larger rates. On the other hand, 
if the rate is very small, the generation of the baryon asymmetry is not efficient from the very beginning and because of that 
the final value is also small. So there is an intermediate magnitude of the rate for which the baryon asymmetry is maximal. 

\begin{figure}[htbp] 
\centering
 \includegraphics[width=.4\textwidth]{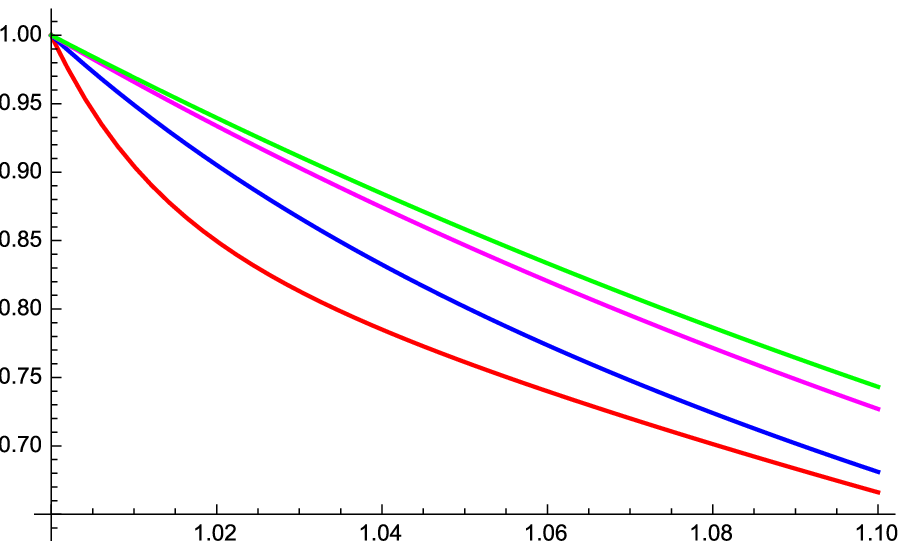} \hspace{.2cm}
\includegraphics[width=.4\textwidth]{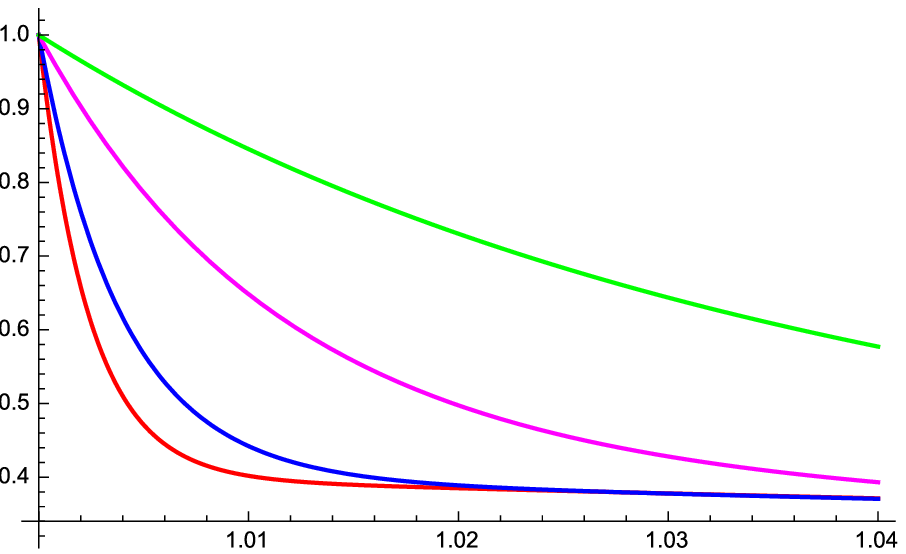}
\caption{
{\it Left}:  Evolution of $\dot \theta (\eta)$, normalized to its initial value, 
for  $ C_B T_ {in}^2 /(5 f^2) =   0.1$ and K = N = 20 (red),  5 (blue),  1 (magenta),  0.3 (green). 
{\it Right}: The same with
$ C_B T_ {in}^2 /(5 f^2) =   1$ and K = N = 50 (red),  30 (blue),  10 (magenta),  3 (green). }
\label{fig-2}       
\end{figure}

The variation of $\dot \theta (\eta)$ calculated according to eq.~(\ref{dot-theta-of-xi}) with $\xi_B(\eta)$ determined from
eq.~(\ref{xi-B-of-eta}) is presented in Fig.~\ref{fig-2}. It is clearly seen, that $\dot \theta$ is not constant, but quite strongly 
changes as a function of temperature or time, especially near the initial moment. 
It means that the basic assumption of the SBG scenario is violated.

\section{Pseudogoldstone case \label{s-psevdo}}

If the potential $U(\theta)$  is non-zero, the equation of motion (\ref{d2-theta-FRW}) cannot be  so easily integrated.
This case  is more efficient for generation of the cosmological baryon asymmetry because the field $\theta (t)$ naturally 
oscillates around the potential minimum, while the mechanism leading to non-zero $\dot\theta$, especially after
inflation, is unclear.
The potential is usually taken in the form:
\be 
U(\theta) =- f^2 m_\theta^2 \cos \theta \rar f^2 m_\theta^2 \theta^2 /2 ,
\label{U-cos}
\ee
where the last equality corresponds to expansion of the cosine near the minimum of the potential. 

To obtain a closed system of equations describing the evolution of $\theta(t)$ with an account of back reaction of the created 
baryons one needs to average the quantum operator $(\dot n_B+ 3 H n_B)$ over the medium. In ref.~\cite{AD-KF} the averaging was performed 
over vacuum state. It corresponds only to decay of $\theta (t)$ while the back reaction of the particles in cosmic plasma
restoring $\theta$-field is neglected.
To include this back reaction we need to use kinetic equation (\ref{kin-eq-gnrl}),   expressing $\dot n_B$
through the collision integral which depends upon $\theta (t)$ and $\xi_B(t)$. As a result a system of the ordinary differential 
and integral equations is obtained which completely determines the evolution of $\theta (t) $ and $n_B (t)$. 
The problem becomes much simpler in thermal equilibrium when the collision  integral is reduced to an 
algebraic relation between $\theta (t)$ and $\xi_B$. However,
this is true only if $\theta $ is slowly varying function of time and $\dot\theta$ is essentially constant. If this is so, we return to the
situation considered in the previous section. The case when the variation of $\theta (t)$ is of importance demands  modification of the 
kinetic equation for the time dependent background, discussed in the following section. 

Note that if the B-nonconserving  reactions are frozen, the baryon
number density remains constant in the comoving volume, i.e. $\dot n_B + 3 H n_B = 0$, so the evolution of $\theta$ is
governed by the free Klein-Gordon equation. Correspondingly  $\theta (t)$ during the equilibrium period simply 
oscillates near the minimum of
the potential with adiabatically decreasing amplitude induced by the cosmological expansion.

In ref.~\cite{spont-BG-1,spont-BG-2} a different approach was taken. It was assumed that the back reaction of the particle 
production on the evolution of $\theta$ could be described by the "friction" term $\Gamma \dot\theta$ which was added to 
the equation of motion:
\be
  f^2 (\partial _t + 3 H  ) \dot\theta + f^2 \Gamma \dot\theta + U'(\theta)   =  0,
  \label{d2-theta-Gamma}
  \ee 
where $\Gamma$ is the rate of B-nonconserving processes.
Comparing this equation with eq. (\ref{d2-theta-FRW}) the authors concluded that $\theta(t)$ oscillates with exponentially
decreasing amplitude, $\sim \exp (-\Gamma t) $ and that 
\be
\dot n_B + 3 H n_B =  f^2\Gamma \dot\theta.
\label{dot-n-B-Gamma}
\ee
However, this might be true only for the decays into empty or overcooled state, as was mentioned in ref.~\cite{spont-BG-2}. 
In this case thermal equilibrium is broken and the identification of $\dot\theta/T$ with $\xi_B$ is questionable. Another problem is
a possibility of description of particle production by $\Gamma \dot\theta$. As it is shown in the paper~\cite{AD-SH}, such
description can only be true, but not necessarily so, for harmonic potential of the field, which produces particles. In the case when 
the interaction is given by $(\dot n_B+ 3 H n_B)$, one has to average this quantum operator over the medium with external field $\theta (t)$.
As a result  a non-local in time expression containing $\theta (t)$ emerges leading to integro-differential equation for $\theta$,
which is not reduced to eq.~(\ref{d2-theta-Gamma}). The problem is treated this way in ref.~\cite{AD-KF}, where the results are
different from those obtained in the papers \cite{spont-BG-1, spont-BG-2}.

\section{Kinetic equation for time-varying amplitude \label{s-kin-eq}}

The canonical kinetic equation (\ref{kin-eq-gnrl}) 
is usually presented for scattering or decay processes in time independent
or slowly varying background with the collision integral giving by eq.~(\ref{I-coll}).

In the case when the interaction proceeds in  time dependent background and/or the time duration of the process is finite, then the energy conservation
delta-function does not emerge and the described approach becomes invalid, so one has to make the time integration with an account of  
time-varying background and integrate over the phase space without energy conservation.

In what follows we consider two-body inelastic process with baryonic number non-conservation with the amplitude obtained from the
last term in Lagrangian (\ref{L-of-theta-1}). At the moment we will not specify the concrete form of the reaction but only will say that it is
the two-body reaction 
\be
a+b \lrar c+d ,
\label{ab-go-cd}
\ee
where $a,b,c$, and $d$ are some quarks and leptons or their antiparticles. The expression for the 
evolution of the baryonic number density, $n_B$, follows from eq. (\ref{kin-eq-gnrl}) after integration of its both sides over $d^3 p_i/(2\pi)^3$.
Thus we obtain:
\be
\dot n_B + 3 H n_B = -\frac{(2\pi)^3}{t_{max}}  \int d\nu_{in} d\nu_{fin}\,\delta({\bf P}_{in} -{\bf P}_{fin} )\, |A|^2 
\left( f_a f_b - f_c f_d \right)
\label{nB-kin-eq}
\ee
where e.g. $d\nu_{in} = {d^3 p_a d^3 p_b}/{[ 4 E_a E_b (2\pi)^6}]$ and
the amplitude of the process is defined as
\be
A = \left(\int_0^{t_{max}} dt\, e^{i [ (E_c+E_d - E_a -E_b)t + \theta (t) ]} \right) F(p_a,p_b,p_c,p_d) ,
\label{amp}
\ee
and $F$ is a function of 4-momenta of the participating particles, determined by the concrete form of the interaction Lagrangian.
In what follows we consider two possibilities: $ F= const$ and $F = \psi^4 \,m_X^{-2}$, where in the last case $\psi^4$ symbolically 
denotes the product of the Dirac spinors of particles $a, b, c$, and $d$.

In the case of equilibrium with respect to baryon conserving reactions the distribution functions have the canonical form
$f_a = \exp (-E_a /T + \xi_a)$, where $\xi_a \equiv \mu_a/T$ is the dimensionless chemical potential. So for constant $F$
the product  $|A|^2 (f_a f_b - f_c f_d) $ depends upon the particle 4-momenta only through $E_{in}$ and $E_{fin}$, where
\be
E_{in} = E_a +E_b,\,\,\,\,{\rm and}\,\,\,\, E_{fin} = E_c+E_d .
\label{E-in-E-fin}
\ee  
Now we can perform almost all (but one)  integrations over the phase space in eq. (\ref{nB-kin-eq}). To this end it is convenient
to change the integration variables, according to:
\be 
\frac{d^3 p_a}{E_a}\,\frac{d^3 p_b}{ E_b } =  d^4 P_{in}\, d^4 R_{in} \,\delta (P^2_{in} + R^2_{in} )\, \delta (P_{in}R_{in}), 
\label{pp-PR}
\ee
where $P_{in} = p_a +p_b$ and $R_{in} = p_a - p_b$ and masses of the particles are taken to be zero. Analogous expressions are
valid for the final state particles. Evidently the time components of the 4-vectors $P$ are the sum of  energies of the incoming and outgoing 
particles, $P^{(4)}_{in} = E_{in} $ and $P^{(4)}_{fin} = E_{fin} $. 

First we integrate over the initial momenta $d^4 P_{in} d^4 R_{in} $ through the  following steps (to avoid an overload of the equations
we skip below the subindex "in" where it is not necessary):\\
1. Integration over $d^3 P_{in}$ (or $d^3 P_{fin}$) with $\delta({\bf P}_{in} - {\bf P}_{fin})$ gives simply 1.\\
2. Taking the integral over 
$ d^4R =  2\pi dR_{4} {\bf R^2} d |{\bf R} |\ d\zeta $
we first integrate over the polar angle using
\be
\delta (P R) = \delta\left (P_4 R_4 - |{\bf P}| |{\bf R}| \zeta \right),
\label{delta-PR}
\ee
so $\zeta = Q_4 R_4/\left( {\bf |R| |Q|}\right)$ and  using the delta-function $\delta (Q_4^2 - {\bf Q}^2 +R_4^2 -  {\bf R}^2 )$
we find that $R_4$ is bounded by $R^2_4 < {\bf Q}^2$, because $|\zeta | <1$. 
The integral over ${\bf R^2} /| {\bf Q}| $ is taken with the written just above delta-function and we are left with the integration
over $dR_4$ in the limits $(-|\bf{Q}|)$ and $(+|\bf{Q}|) $.  So the  integration over the  initial momenta is reduced finally to 
$2 \pi dQ_4 $. \\
3. Proceeding along the same lines with the  integration over the phase volume of the final particles, but without $\delta({\bf P}_{in} -{\bf P}_{fin})$ 
we obtain:
\be
(2\pi)^3\,\int d\nu_{in} d\nu_{fin} \delta({\bf P}_{in} -{\bf P}_{fin})= \frac{1}{2^9 \pi^6} \int dE_{in}dE_{fin} \,d |{\bf Q}_{fin}|\, | {\bf Q}_{fin}^2| .
\label{dE-dE-dQ}
\ee
Naively we should expect that  the
integration over $|{\bf Q}_{fin} | $ lays in the limits from 0 to $E_{fin}$ because 
\be
{\bf Q}_{fin}^2 = E_c^2 + E_d^2 + 2 E_c E_d \zeta < (E_c+E_d)^2 = E_{fin}^2 , 
\label{lim-Q}
\ee
but there is a constraint ${\bf Q}_{fin} = {\bf Q}_{in}$, so the upper limit on $ |{\bf Q}_{fin}|$ is the smaller out of $E_{fin}$ and
$E_{in}$. Let us introduce new notations: $E_+ = E_{in}+E_{fin} $ and $E_- = E_{in} - E_{fin} $. 
 It is easy to check that $E_{fin} > E_{in} $ for $E_- <0$ and   $E_{fin} < E_{in} $ for $E_-  >0$. 
Thus for $E_- <0$ the
integration over  $d |{\bf Q}_{fin}|$ in eq.~(\ref{dE-dE-dQ})  gives   $ E_{in}^3/3 $, while for $E_- >0$ the result is $ E_{fin}^3/3 $. \\
4. So we are left with the integral over $dE_{in} dE_{fin}$ which is convenient to rewrite as
\be
\int dE_{in} dE_{fin} = dE_+ \,dE_-/2,
\label{dE+dE-}
\ee
Note that the amplitude A (\ref{amp}) depends only on $E_-$ but not on $E_+$,
while the products of the particle densities in the phase space are
\be
f_a f_b = \exp \left( -\frac{E_+ + E_-}{2T} + \xi_a + \xi_b \right) \,\,\, {\rm and}\,\,\, f_c f_d = \exp \left( -\frac{E_+ - E_-}{2T} + \xi_c + \xi_d \right). 
\label{prod-f}
\ee
5. The integral over $dE_+$ can be taken explicitly but first we need to establish the integration limits.
The original integration over $dE_{in} dE_{fin}$ is taken from 0 to $\infty$, so the integral over $dE_+$ runs from $|E_-|$ to $\infty$ and
the integral over $dE_-$ runs from $(-\infty)$ to $(+\infty)$. It is convenient to separate the integration over $dE_+$ into two parts for positive and
negative $E_-$. For positive $E_-$ we find
\be
&&\int_{E_-}^{\infty}  dE_+  \left( \frac{E_+ - E_-}{2} \right)^3  \exp \left( -\frac{E_+  +  E_-}{2T} \right)  = 
12 T^4 e^{- y} , \nonumber \\
&&\int_{E_-}^{\infty}  dE_+  \left( \frac{E_+  -  E_-}{2} \right)^3  \exp \left( -\frac{E_+  -  E_-}{2T} \right)  = 12 T^4,
\label{int-pos-E}
\ee
where $y=E_-/T$.
For negative $E_-$ we obtain the same results with an interchange of the initial and final states, i.e. $f_a f_b \lrar f_c f_d$ 
and with $y \rar |y|$. Effectively it corresponds to the change of sign of $\theta (t)$ in eq.~(\ref{amp}).

Thus, collecting all  the factors (\ref{prod-f}), we finally obtain:
\be
\dot n_B + 3 H n_B = - \frac{T^5}{2^5 \pi^6\, t_{max}}\, \int_0^\infty dy  \left[ e^{\xi_a + \xi_b} \left(  |A_+|^2 + |A_-|^2 e^{-y}   \right) -
e^{\xi_c+ \xi_d} \left(  |A_{ - }|^2 + |A_{+}|^2 e^{-y}   \right) 
\right] ,
\label{dot-nB-gen}
\ee
where $A_+$ is the amplitude taken at positive $E_-$, while $A_-$ is taken at negative $E_-$.  With the substitution 
$E_- \rar |E_-|$ the only difference between $A_+$ and $A_-$ is that  $A_- (\theta) = A_+ (-\theta)$. 

The equilibrium is achieved when the integral in eq. (\ref{dot-nB-gen}) vanishes.
  This point determines the equilibrium values of the chemical 
potentials in external $\dot\theta $ field.
Clearly it takes place at:
\be
\xi_a+\xi_b -\xi_c -\xi_d = \frac{ \langle |A_+|^2 e^{-y} + |A_-|^2 \rangle } {  \langle |A_+|^2  + |A_-|^2 e^{-y} \rangle} -1 ,
\label{equiv-sol}
\ee
where the angular brackets mean integration over $dy$ as indicated in eq. (\ref{dot-nB-gen}).

This results above are obtained for the amplitude which does not depend upon the participating particle momenta. 
The calculations would be somewhat more complicated if this restriction is not true. For example if the baryon 
non-conservation takes place in  four-fermion interactions, then the amplitude squared can contain the terms of 
the form $(p_a p_b)^2 /m_X^4$ or  $(p_a p_c)^2 /m_X^4$, etc. The effect of such terms results in a 
change of the numerical coefficient in eq. (\ref{dot-nB-2}) but  the latter is unknown anyhow,
and what is more important the temperature coefficient in front of  the integral in this equation would change from $T^5$
to $T^9/m_X^4$.

\section{Examples of time-varying $\theta$ \label{s-examples}}

\subsection{Constant $\dot \theta$ \label{ss-const-dot-tjeta}}

This is the case usually considered in the literature and the simplest one. The integral (\ref{amp}) is taken analytically resulting in:
\be
|A|^2 \sim \frac{ 2- 2\cos[ ( \dot\theta - E_-) t_{max}]}{ (\dot\theta - E_-)^2} .
\label{A-const-dot-theta}
\ee
Here $E_-$ is running over the positive semi-axis, see eq.~(\ref{int-pos-E}) and  comments around it.

For large $t_{max}$ this expression tends to $\delta (E_--\dot\theta)$, so $ |A_+|^2 = 2\pi \delta (E_-  - \dot\theta) t_{max} $ 
and $ |A_-|^2 = 2\pi \delta (E_-  +\dot\theta) t_{max} =0$, if $\dot\theta >0$ and vice versa otherwise.
Hence the equilibrium solution is
\be
\xi_a+\xi_b -\xi_c -\xi_d - \dot\theta /T= 0, 
\label{equil-1}
\ee
coinciding with the standard result. 

The limit of $\dot\theta = const$ corresponds to the energy non-conservation by
the rise (or drop) of the energy of the final state in reaction (\ref{ab-go-cd}) exactly by $\dot\theta$.
However if  $t_{max}$  is not sufficiently large, the non-conservation of energy is not equal to $\dot\theta$ 
but somewhat spread out and the
equilibrium solution would be different. There is no simple analytical expression in this case, so 
 we have to take the integrals over $y$ in eq.~(\ref{equiv-sol})  numerically. 
 
\begin{figure}[htbp] 
\centering
 \includegraphics[width=.4\textwidth]{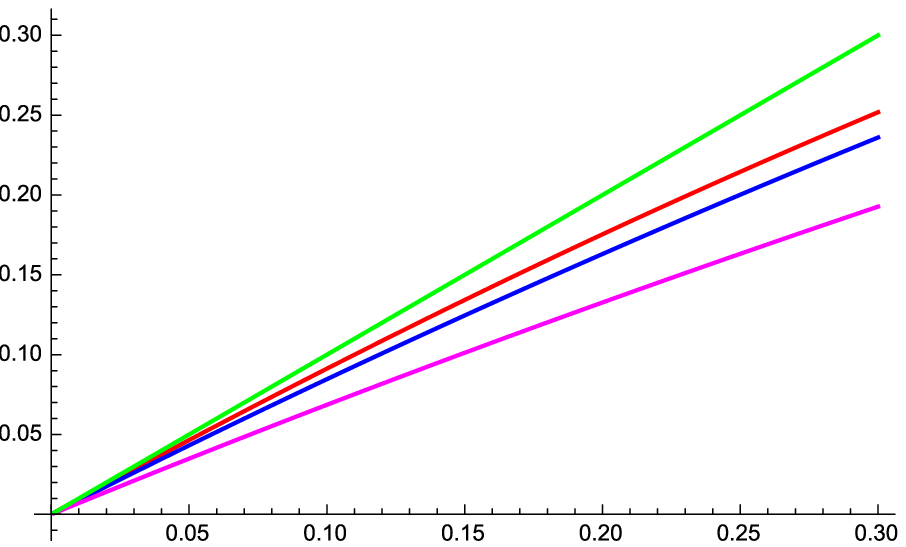} \hspace{.5cm}
\includegraphics[width=.4\textwidth]{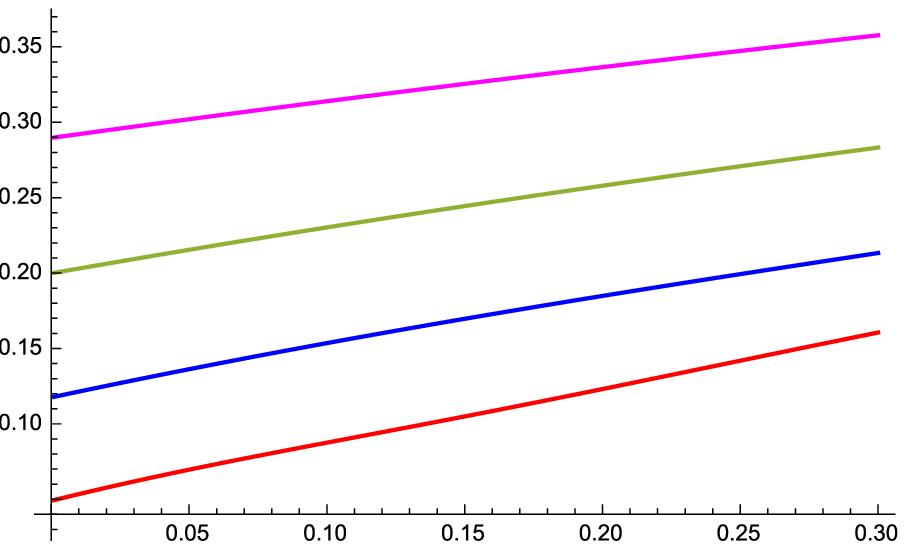}
\caption{
{\it Left}: The r.h.s. of eq.~(\ref{equiv-sol})
 as a function of  $\dot\theta/T$ for  the cut-off of the time integration in eq.~(\ref{A-const-dot-theta}):
$\tau \equiv t_{max} T =$ 30 (red); 10 (blue); 3 (magenta) and $\dot\theta/T$ (green).
{\it Right}: The relative difference between  the  r.h.s. of eq.~(\ref{equiv-sol}) and $\dot\theta/T$, normalized to $\dot\theta/T$,
as a function of $\dot\theta$ for
$ \tau =$ 30 (red), 10 (blue), 5 (green), 3 (magenta).}
\label{fig-3}       
\end{figure}

The results of the calculations are presented in Fig~\ref{fig-3}. In the left panel  the values of the r.h.s. of eq.~(\ref{equiv-sol})
are presented as a function of  $\dot\theta/T$ for  the cut-off of the time integration in eq.~(\ref{A-const-dot-theta}) equal to:
$\tau \equiv t_{max} T = 30; 10; 3$.   
The larger is the integration time the closer are the lines to $\dot\theta/T$, which is also depicted.

In the right panel the relative differences between  the  r.h.s. of eq.~(\ref{equiv-sol}) and $\dot\theta/T$, normalized to $\dot\theta/T$,
as a function of $\dot\theta/T$ for different
maximum time of the integration are presented. 
We see that for $\tau = 30$ the deviations are less than 10\%,
while for $\tau = 3$ the deviations are about 30\%. If we take $\tau$ close to unity, the deviations are about 100\%. 
The value of $\dot\theta/T$ is bounded from above by approximately 0.3 because at large $\dot\theta/T$ the linear expansion, used
in our estimates, is invalid.
 
The realistic values of $\tau$ depend upon the model parameters. There is one evident limit related to the cosmological
expansion, which implies $\tau < t_{cosm} T \sim T/H \sim m_{Pl} /T $. Here $m_{Pl}$ is the Planck mass,
$H$ is the Hubble parameter, and $T_{cosm} \sim 1/H$, so the effects of the expansion may be significant only near 
the Planck temperature. Another upper bound on $\tau$ is presented by the kinetic equations which demands 
the characteristic time variation to be close (at least initially) to the inverse reaction rate $\gamma \sim T^5/m_X^4$. The discussed
effects would have an essential impact on the approach to equilibrium  for $T\sim m_X$ which might be
realistic.

\subsection{Second order Taylor expansion of $\theta (t)$ \label{ss-Taylor}}

As we have seen in the previous subsection the approximation $\dot \theta = const$ is noticeably violated. 
Here we assume that $\theta (t)$ can be approximated as 
\be
\theta (t) = \dot \theta \,t + \ddot \theta\, t^2/2,
\label{theta-Taylor}
\ee
where $\dot\theta$ and $\ddot\theta$ are supposed to be constant or slowly varying. 
In this case the integral over time (\ref{amp}) can also be taken analytically but the result is rather complicated.
We need to take the integral
\be
\int_0^{t_{max}} dt \exp [i \theta (t) ] .
\label{int-theta-of-t}
\ee
Its real and imaginary parts are easily expressed though the Fresnel functions.  So the amplitude squared is given by the
functions tabulated in Mathematica and the position  of the equilibrium point can be calculated, as in the previous case, 
by numerical calculation of one dimensional integral.

\begin{figure}[htbp] 
\centering
 \includegraphics[width=.4\textwidth]{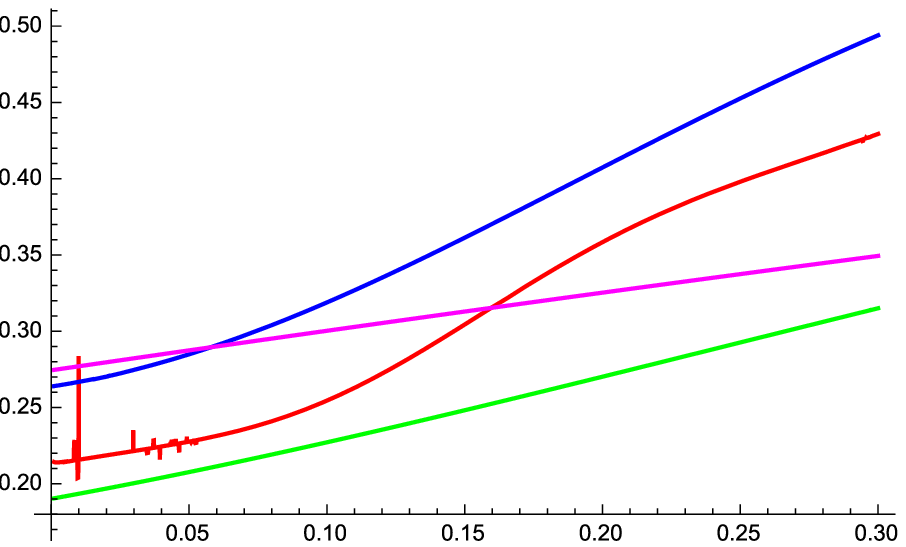} \hspace{.5cm}
\includegraphics[width=.4\textwidth]{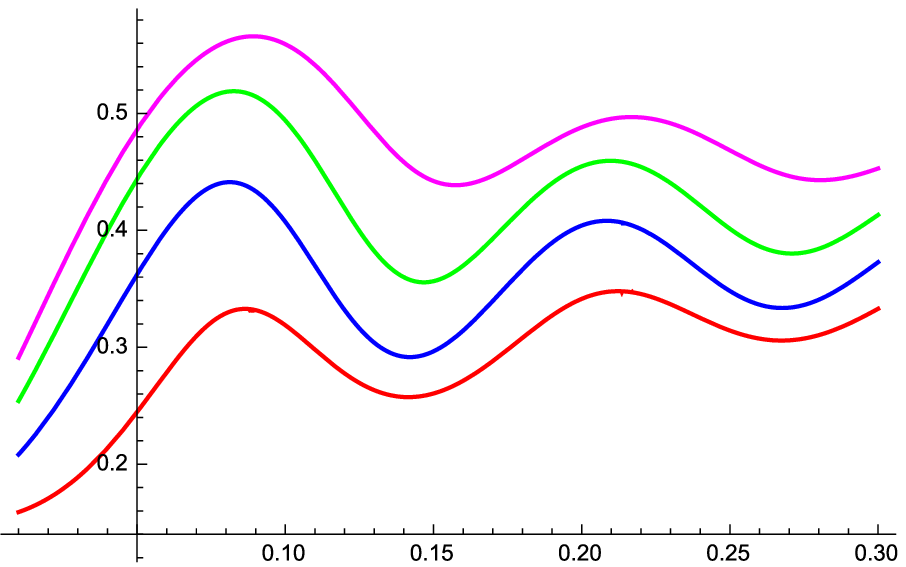}
\caption{
{\it Left}: The relative difference between  the  r.h.s. of eq.~(\ref{equiv-sol}) and $\dot\theta/T$, normalized to $\dot\theta/T$,
 as a function of  $\dot\theta/T$ for  the cut-off of the time integration: 
 $\tau = $ 30 (red), 10 (blue), 5 (green), 3 (magenta) for fixed $\ddot \theta/T^2 =  0.1 $.
{\it Right}: 
The same difference
as a function of $\ddot\theta /T^2$ for  fixed time of integration $\tau = 10 $ and different  
$\dot \theta/T = $ 0.1 (red), 0.2 (blue), 0.3 (green), 0.4 (magenta).}
\label{fig-4}       
\end{figure}

The r.h.s. of eq.~(\ref{equiv-sol}) as a function of $\dot\theta /T$ for different values of $\tau$ is presented in Fig.~\ref{fig-4}, at the left panel. 
It is interesting that the dependence on $\tau$ is non-monotonic. This may be understood by diminishing of the impact 
of $\ddot\theta t^2$ at smaller time interval.

To check the dependence on $\ddot\theta$ we calculated again the  r.h.s. of eq.~(\ref{equiv-sol}) but now as a function of 
$\ddot\theta/T^2$ presented at the right panel of Fig.~\ref{fig-4} for fixed time of integration and different values of $\dot \theta /T$. 
We see that the equilibrium point oscillates as a function $\ddot\theta$.

\subsection{Oscillating $\theta(t)$ \label{ss-osc-theta} }

\begin{figure}[htbp] 
\centering
 \includegraphics[width=.4\textwidth]{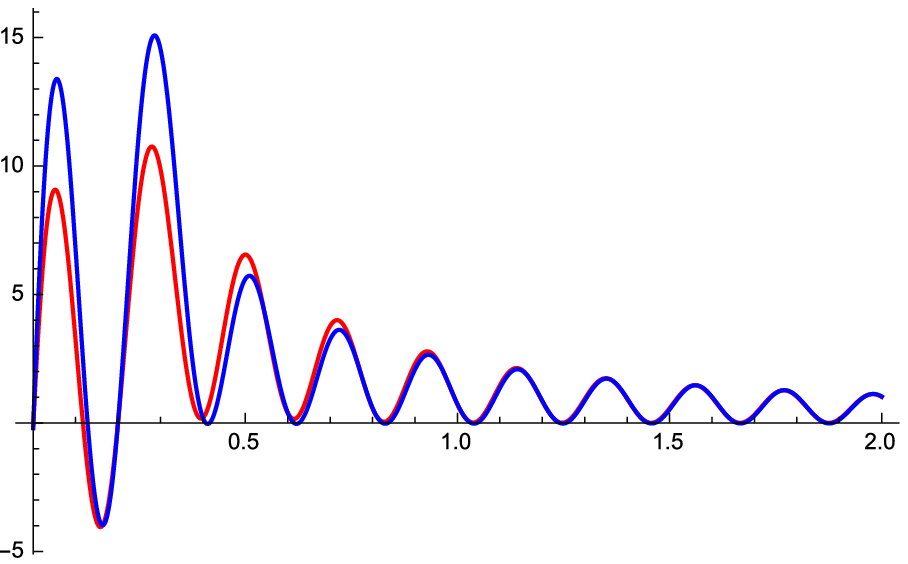} \hspace{.5cm}
\includegraphics[width=.4\textwidth]{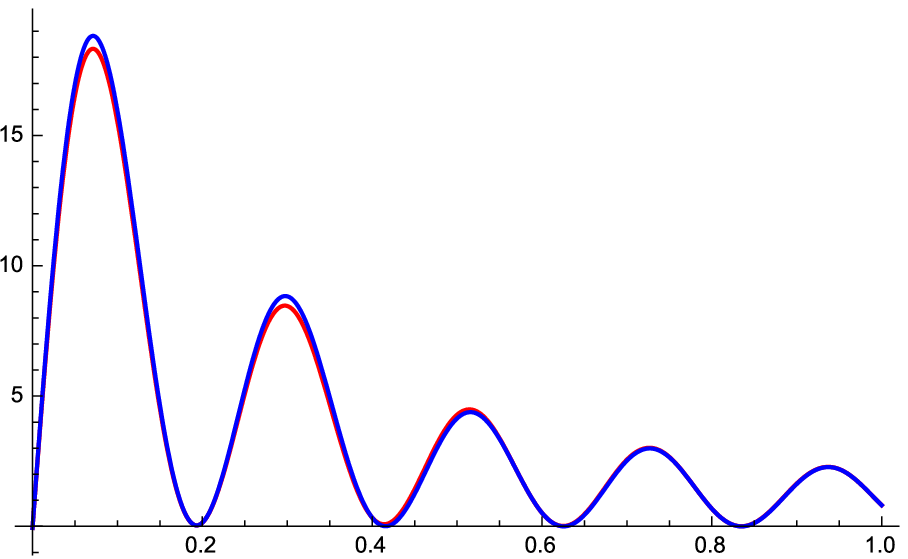}
\caption{  Exact (red) and approximate (blue) expressions for the amplitude (\ref{amp}) with (\ref{expand-exp}) as functions of $mt$ for 
$\theta_0=1$ ({\it left}) and $\theta_0=0.3$ ({\it right}).}
\label{fig-5}       
\end{figure}

If the potential of $\theta$ is non-vanishing,  its evolution would be more complicated. The potential $U(\theta)$
should be a periodic function of the angle $\theta$ and so it is often taken as $ m^2 \cos \theta $. We assume that the
field $\theta $ is initially near the minimum of the potential, which in this case can be approximated as $U = m^2 \theta^2/2 $, 
where $m$ is the mass of the theta-field.
 In absence of back reaction of the produced  baryons $\theta (t)$ should evolve as
\be
\theta (t) = \theta_0 \cos (m t +\phi ) .
\label{theta-harm}
\ee
Unfortunately the integral (\ref{amp}) cannot be taken analytically and the numerical calculations with 2-dimensional
integrals are quite time consuming. However, the integrand can be expanded as 
\be
e^{i\theta (t)}  = 1 + i \theta_0  \cos (m t +\phi ) .
\label{expand-exp}
\ee
In this approximation the integral  (\ref{amp}) can be easily taken analytically. Thus also in this case we can reduce
the calculation of the deviation of  the algebraic sum of dimensionless chemical potentials from $\dot \theta/T$ (\ref{equil-1}) 
to the numerical calculation of one dimensional integral. However, to be sure in the safely of the procedure
it is desirable to compare the time integrated exact amplitude with the approximate expanded one. Numerical comparison
shows indeed that even for $\theta_0 = 1$ the corrections are negligible, while for $\theta_0  \leq 0.5$ they are practically  
indistinguishable (see Fig.~\ref{fig-5}).

\begin{figure}[htbp] 
\centering
 \includegraphics[width=.37\textwidth]{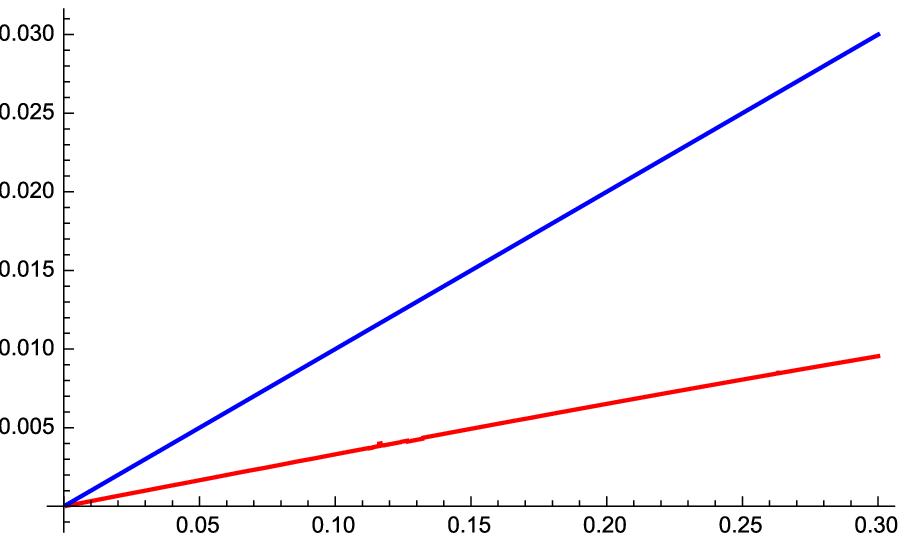} \hspace{.5cm}
\includegraphics[width=.37\textwidth]{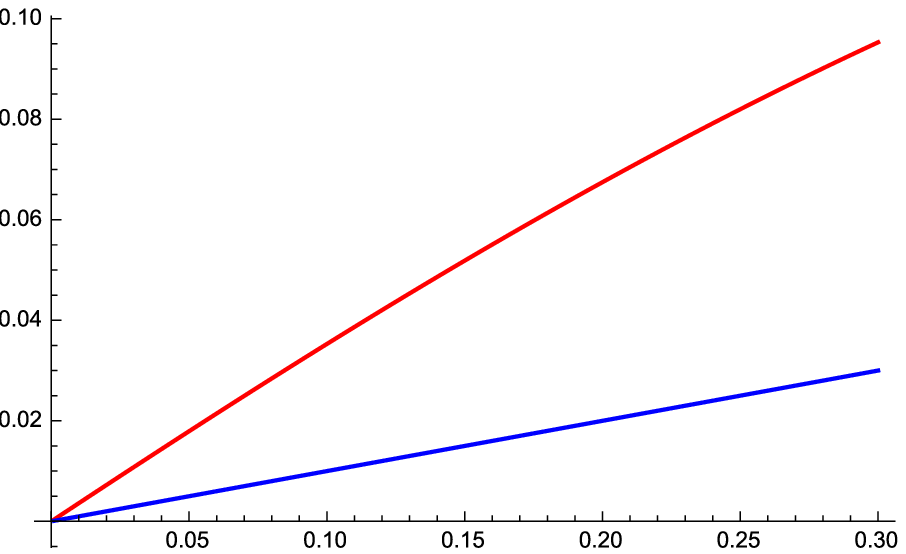}
\caption{  
{\it Left}:
Red curve: the  r.h.s. of eq.~(\ref{equiv-sol}) for $m/T = 0.1$   and the maximal time of interation $\tau = 30$,
blue curve: $\dot\theta/T$ as  functions of $\theta_0$, see eq.~(\ref{theta-harm}). 
{\it Right}: the same with the maximal time of integration $\tau = 3$.
}
\label{fig-6}       
\end{figure}

The deviation of the r.h.s. of eq. (\ref{equiv-sol}) from $\dot\theta/T$ is demonstrated in Fig.~\ref{fig-6}. The difference with the standard
predictions of SBG can be significant if the mass of $\theta $ is not negligible, so the oscillations of $\theta$ manifest themselves
during "time" $\tau$. So the standard SBG, for which the baryonic chemical potential  is proportional to $\dot \theta $, 
is not accurate at large times or, better to say, for large $m t _{max}$. On the other
hand, as we see in these figures, for small $\tau$ the deviations are also quite noticeable, but now the effect is
related to the energy spread because of the finite time integration. As it  is seen in the figures, the effect changes sign - the relative
positions of red and blue curves interchange.

\section{Conclusion \label{s-conclude}}

To summarize, we have clarified the relation between Lagrangian and Hamiltonian in SBG scenario. We argue that in the standard
description $\dot\theta$ is not formally the chemical potential, though in thermal equilibrium $\dot \theta$ may tend to 
the chemical potential  with the numerical coefficient
which depends upon the model. However,  this result is not always true but depends upon the chosen representation of the "quark"
fields. In the theory described by the Lagrangian (\ref{L-of-theta-1}) which appears "immediately" after the spontaneous symmetry 
breaking,  $\theta (t)$ directly enters the interaction term and in equilibrium  $\mu_B \sim \dot\theta$ indeed. On
the other hand, if we transform the quark field, so that the dependence on $\theta$ is shifted to the bilinear product of the quark 
fields (\ref{L-of-theta-2}), then chemical potential in equilibrium does not tend to $\dot\theta$, but  to zero. Still, the 
magnitude of the baryon asymmetry in equilibrium is always proportional to $\dot\theta$. 

It can be seen, according to the equation
of motion of the Goldstone field that $\dot\theta/T$ drops down
in the course of the cosmological cooling as $T^2$, so the baryon number density in the comoving volume decreases in the
same way. So to avoid the complete vanishing of $n_B$ the baryo-violating interaction should switch-off at some non-zero 
and not very small temperature.  The dependence of the baryon asymmetry on the interaction strength is non-monotonic. 
Too strong and too weak interactions lead to 
small baryon asymmetry, as is presented in Fig.~\ref{fig-1}. 

The assumption of a constant or slowly varying $\dot\theta$, which is usually done in the SBG scenario, may be not fulfilled and
to include the effects of an arbitrary variation of $\theta (t)$, as well as the effects of the finite time integration, we transformed the kinetic 
equation in such a way that it becomes operative in non-stationary background. A shift of the equilibrium value of the baryonic
chemical potential due to this effect is numerically calculated. 

In spite of these corrections to the standard SBG scenario, it
remains a viable mechanism for creation of the observed cosmological excess of matter over antimatter. 
However,  this mechanism is not particularly efficient in the case of pure 
spontaneous symmetry breaking, when the potential of the $\theta$-field is absent. 
 Non-zero potential $U(\theta)$, which can appear as a result of an
explicit breaking of the baryonic $U(1)$-symmetry in addition to the spontaneous breaking may grossly enhance the efficiency of 
the spontaneous baryogenesis.
The evaluation of the efficiency demands numerical solution of the ordinary differential equation of motion for the $\theta$-field together
with the integral  kinetic equation. In the case of thermal equilibrium the  kinetic equation is reduced to an algebraic one and the system
is trivially investigated. The out-of-equilibrium situation is much more complicated technically and will be studied elsewhere.

We assumed that the symmetry breaking phase transition in the early universe occurred instantly. It may be a reasonable approximation,
but still the corrections can be significant. This can be also a subject of future work.

There remains the problem of the proper definition of the fermionic Hamiltonian but presumably it does not have an important impact on the
considered here problems and thus is neglected.
\\[3mm]
{\bf Acknowledgement}
We thank A.I. Vainshtein for stimulating criticism.  The work of E.A. and A.D. was supported by the RNF Grant N 16-12-10037. 
V.N. thanks the support of the Grant RFBR 16-02-00342.

\end{document}